\documentclass[a4paper,amsmath,amssymb,nofootinbib,showpacs,notitlepage,byrevtex,showkeys,prd
]{revtex4-1}

\usepackage{hyperref}
\usepackage{graphicx}

\newcommand*{\be}{\begin{equation}}
\newcommand*{\ee}{\end{equation}}
\newcommand*{\vev}[1]{\left\langle #1 \right\rangle}
\DeclareMathOperator{\Det}{Det}	
\newcommand*{\coloneq}{\mathrel{\mathop:}=}
\newcommand*{\eqcolon}{=\mathrel{\mathop:}}
\newcommand*{\Eqref}[1]{Eq.~\eqref{#1}}
\renewcommand{\vec}[1]{\boldsymbol{\mathrm{#1}}}
\newcommand*{\vp}{\vec{p}}
\newcommand*{\vq}{\vec{q}}
\newcommand*{\vk}{\vec{k}}
\newcommand*{\vx}{\vec{x}}
\newcommand*{\vy}{\vec{y}}
\newcommand*{\vpn}{\vec{p_0}}
\newcommand{\nn}{\nonumber}
\newcommand{\nnnl}{\nonumber\\}	
\newcommand*{\fref}[1]{Fig.~\ref{#1}}
\newcommand*{\calD}{\mathcal{D}}
\newcommand*{\calJ}{\mathcal{J}}
\renewcommand*{\d}[1][]{\mathop{}\!\mathrm{d}^{#1}}
\newcommand*{\abs}[1]{\ensuremath{\lvert#1\rvert}}
\newcommand*{\e}{\ensuremath{\mathrm{e}}}
\newcommand*{\deltabar}{\delta\mkern-8mu\mathchar'26}
\makeatletter
\def\I{\@ifstar\@ImU\@@ImU}
\newcommand*{\@ImU}{\ensuremath{{\mathrm{i}\mkern1mu}}}
\newcommand*{\@@ImU}{{\ensuremath{\mathrm{i}}}}
\newcommand*{\sigmacoul}{\ensuremath{\sigma_\mathrm{C}\@ifnextchar{^}{}{^{}}}} 
\newcommand*{\Nc}{\@ifnextchar{^}{\mathrlap{N_\mathrm{c}}\phantom{N}}{N_\mathrm{c}}}
\makeatother

\allowdisplaybreaks 

\begin{document}

\title{Vertex functions of Coulomb gauge Yang--Mills theory}

\author{Markus Q.~Huber}
\affiliation{Institute of Physics, University of Graz, Universit\"atsplatz 5, 8010 Graz, Austria}
\author{Davide R.~Campagnari}
\author{Hugo Reinhardt}
\affiliation{Institut f\"ur Theoretische Physik, Eberhard-Karls-Universit\"at T\"ubingen,
Auf der Morgenstelle 14, 72076 T\"ubingen, Germany}
\date{\today}

\begin{abstract}
The canonical recursive Dyson--Schwinger equations for the three-gluon and ghost-gluon vertices
are solved numerically. The employed truncation includes several previously neglected diagrams
and includes back-coupling effects.
We find an infrared finite ghost-gluon vertex and an infrared diverging three-gluon vertex.
We also compare our results with those obtained in previous calculations, where
bare vertices were used in the loop diagrams.
\end{abstract}

\pacs{11.10.Ef, 12.38.Aw, 12.38.Lg}

\keywords{Hamiltonian approach, ghost-gluon vertex, three-gluon vertex, Coulomb gauge}

\maketitle


\section{Introduction}

In recent years many efforts have been undertaken to develop non-perturbative approaches
to continuum Quantum Chromodynamics (QCD). Among these are variational approaches to Yang--Mills theory in Coulomb 
gauge which use Gaussian trial ans\"atze for the Yang--Mills vacuum wave functional \cite{Schutte:1985sd,Szczepaniak:2001rg,Feuchter:2004mk}.
The approach of Ref.~\cite{Feuchter:2004mk} has given a decent description
of the infrared sector of the theory yielding, among other things, an infrared divergent
gluon energy \cite{Epple:2006hv}, a perimeter law for the 't Hooft loop \cite{Reinhardt:2007wh} (both are manifestations of
confinement), a color dielectric function of the Yang--Mills vacuum in accord with the dual superconductor picture of the QCD vacuum \cite{Reinhardt:2008ek}, and a critical temperature of the 
deconfinement phase transition in the right ballpark (of about 275 MeV) \cite{Reinhardt:2011hq,*Heffner:2012sx, Reinhardt:2012qe,*Reinhardt:2013iia}.
Furthermore, the obtained static gluon propagator is in satisfactory agreement 
with the lattice data \cite{Burgio:2008jr}, both in the infrared and in the ultraviolet,
but misses some strength in the mid-momentum regime. Preliminary studies of Ref.~\cite{Campagnari:2010wc}
show that the missing strength can be attributed to the absence of non-Gaussian terms
in the trial Yang--Mills vacuum wave functional ignored in previous considerations.

In Ref.~\cite{Campagnari:2010wc} a general variational approach to quantum field theories was developed, which is capable of 
using non-Gaussian trial wave functionals. The crucial point in this approach was to realize that once  the vacuum wave 
functional is written as the exponential of  some  action functional given by polynomials of the fields whose 
coefficients  are  treated as variational kernels, one can exploit Dyson--Schwinger equation techniques 
to express the various vacuum expectation values of the fields (viz.~propagators and vertices) and, in particular, the vacuum expectation
value of the Hamiltonian in terms of the 
variational kernels. In this way the variational approach can be carried out for non-Gaussian vacuum wave functionals.
In Ref.~\cite{Campagnari:2010wc} the approach was worked out for pure Yang--Mills theory using an ansatz for the vacuum wave 
functional which contains up to fourth-order polynomials in the gauge field, see Eqs.~\eqref{mad2} and \eqref{mad4} below. In particular,
the corresponding Dyson--Schwinger equations for the propagators and leading vertices were derived. In the present
paper we solve the resulting Dyson--Schwinger equations for the ghost-gluon and three-gluon vertices.

The organization of the paper is as follows: In Sec.~2 we briefly review the essential ingredients of the approach of 
Ref.~\cite{Campagnari:2010wc}. In Sec.~3  we present the Dyson-Schwinger equations for the ghost-gluon and three-gluon vertices. The 
numerical solutions of these equations are presented in Sec.~4. Our conclusions are given in Sec.~5. The Appendix contains some explicit expressions for the integral kernels. 


\section{Hamiltonian Approach to Yang--Mills Theory}

The Hamiltonian approach to Yang--Mills theory rests upon the canonically
quantized theory in the temporal (Weyl) gauge, $A_0^a=0$. 
As a consequence of this gauge, Gauss's law does not show up in the Heisenberg equations
of motion but has to be imposed as a constraint on the wave functional, which in the absence of
matter fields guarantees its gauge invariance. Furthermore, this gauge does not fix the gauge completely
but still leaves invariance with respect to time-independent gauge transformations.
Fixing this residual gauge invariance by imposing the Coulomb gauge $\partial_i A_i^a=0$
one can explicitly resolve Gauss's law for the longitudinal part of the momentum operator.
The longitudinal part of the kinetic energy results then in an extra term in the Hamiltonian,
the so-called Coulomb Hamiltonian, mediating a two-body interaction between colour charges.
One ends with a theory defined entirely in terms of the transverse gauge field. In this
theory the vacuum expectation
value (VEV) of an operator $K[A]$ depending on the transverse gauge field $A$ is given by
\be\label{mad1}
\vev{K[A]} = \int \calD A \: \calJ_A \: \abs{\varPsi[A]}^2 \: K[A] ,
\ee
where $\varPsi[A]$ is the vacuum wave functional, and
$\calJ_A=\Det(G_A^{-1})$ is the Faddeev--Popov determinant of Coulomb gauge with
\be\label{ggvdse0}
G_A^{-1}{}^{ab}(\vx,\vy) = \bigl( - \delta^{ab} \partial^2 - g f^{acb} A_i^c(\vx) \partial_i \bigr) \delta(\vx-\vy)
\ee
being the Faddeev--Popov operator. In \Eqref{ggvdse0} $g$ is the coupling constant
and $f^{acb}$ are the structure constants of the $\mathfrak{su}(\Nc)$ algebra.
The functional integration in \Eqref{mad1} runs over transverse field configurations
$\partial_i A_i^a=0$ and is, strictly speaking, restricted to the first Gribov region.

In the following we use a compact notation in which a numerical index stands for the continuous
spatial coordinate as well as for the discrete indices (colour and, possibly, Lorentz), e.g.~$A(1)\equiv A^{a_1}_{i_1}(\vx_1)$.
A repeated label implies summation over
the discrete indices and integration over the coordinates.

In this work we focus our attention on the Yang--Mills three-point functions, namely the
ghost-gluon and the three-gluon vertex. The full ghost-gluon vertex $\widetilde\Gamma$ is defined by
\be\label{ggv-def}
\vev{G_A(1,2) \, A(3)} \eqcolon - \widetilde{\Gamma}(1',2';3') \, G(1,1') \, G(2',2) \, D(3',3) .
\ee
Here, $G_A$ is the
inverse Faddeev--Popov operator [see \Eqref{ggvdse0}], and $G$ and $D$ are, respectively, the ghost propagator
\be\label{ghp-def}
G(1,2) \eqcolon \vev{G_A(1,2)}
\ee
and the gluon propagator
\[
D(1,2) \eqcolon \vev{A(1) A(2)} .
\]
Similarly we define the three-gluon vertex $\Gamma_3$ by
\be\label{3gv-def}
\vev{A(1) A(2) A(3)} = - \Gamma(1',2',3') \, D(1',1) \, D(2',2) \, D(3',3) .
\ee

The vacuum wave functional has in principle to be found by solving the (functional)
Schr\"odinger equation, which of course cannot be done rigorously in $3+1$ dimensions.%
\footnote{In $1+1$ dimensions the Schr\"odinger equation can be solved exactly \cite{Reinhardt:2008ij}.}
Writing the square modulus of the vacuum wave functional as
\be\label{mad2}
\abs{\varPsi[A]}^2 \eqcolon \exp\{-S[A]\} ,
\ee
\Eqref{mad1} is formally equivalent to a Euclidean field theory described by an ``action'' $S[A]$.
We can exploit this equivalence to derive Dyson--Schwinger-type equations, which
allow us to relate the various $n$-point functions to the variational kernels of the
vacuum wave functional, i.e.~of the action $S[A]$. These equations are derived from the expectation
values \Eqref{mad1} of the canonical theory in a recursive way and will hence be referred
to as canonical recursive Dyson--Schwinger equations (CRDSEs)
To derive these equations we start from the functional identity
\be\label{mad3}
0 = \int \calD A \: \frac{\delta}{\delta A} \bigl\{ \calJ_A \: \e^{-S[A]} \: K[A] \bigr\} .
\ee
The ``action'' $S[A]$ defines the trial ansatz for our vacuum wave functional.
In Ref.~\cite{Campagnari:2010wc} an ansatz of the form
\be\label{mad4}
S[A] = \omega A^2 + \frac{1}{3!} \: \gamma_3 \, A^3 + \frac{1}{4!} \: \gamma_4 \, A^4
\ee
was considered, where $\omega$, $\gamma_3$, and $\gamma_4$ are variational kernels to be
determined by minimization of the vacuum energy.
With this ansatz the CRDSEs derived from \Eqref{mad3} resemble the usual
DSEs of Landau gauge Yang--Mills theory in $d=3$ dimensions with the bare
vertices of the usual Yang--Mills action replaced by the variational kernels.

The CRDSEs are not equations of motion in the usual sense, but rather relations between the
Green functions and the (so far undetermined) variational kernels. In fact, the CRDSEs are
needed when non-Gaussian trial wave functionals are used in order to express the various
correlation functions, and in particular the vacuum energy density, in terms of the variational kernels.

\begin{figure}
\centering
\includegraphics{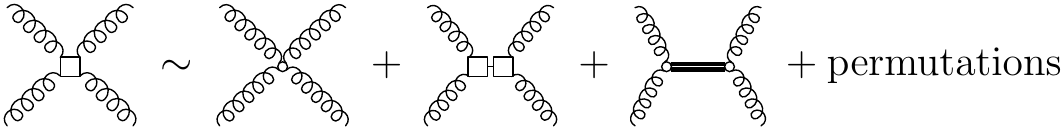}
\caption{\label{fig:4gk}Schematic representation of the four-gluon variational kernel
$\gamma_4$ [Eq.~\protect\eqref{k4}]. Empty boxes represent variational kernels, while
the small empty dot stands for the trivial tensor structure Eq.~\protect\eqref{t4}. The
double line represents the Coulomb propagator Eq.~\protect\eqref{eq:F}.}
\end{figure}

In Ref.~\cite{Campagnari:2010wc} the CRDSEs were used to calculate the VEV
of the Hamiltonian in
the vacuum state defined by Eqs.~\eqref{mad2} and \eqref{mad4}, resulting in an energy
functional
\[
\vev{H_\mathrm{YM}} = E[\omega,\gamma_3,\gamma_4] .
\]
By using a skeleton expansion, the vacuum energy can be expanded at the desired order of
loops. In Ref.~\cite{Campagnari:2010wc} the vacuum energy was calculated up to two-loop order.
Extremizing the vacuum energy density with respect to $\gamma_3$ and $\gamma_4$ results in the following
equations for the three- and four-gluon variational kernels \cite{Campagnari:2010wc}:
\be\label{k3}
\gamma_{ijk}^{abc}(\vp,\vq,\vk) = \frac{2\,g\, T^{abc}_{ijk}(\vp,\vq,\vk)}{\Omega(\vp)+\Omega(\vq)+\Omega(\vk)}
\ee
and
\be\label{k4}
\begin{split}
\bigl[ \Omega(\vk_1) &+ \Omega(\vk_2) + \Omega(\vk_3) + \Omega(\vk_4) \bigr] \, \gamma^{abcd}_{ijkl}(\vk_1,\vk_2,\vk_3,\vk_4) =
2 \, g^2 \, T^{abcd}_{ijkl} \\
-\frac12 &\biggl\{
\gamma^{abe}_{ijm}(\vk_1,\vk_2,-\vk_1-\vk_2) \, t_{mn}(\vk_1+\vk_2) \, \gamma^{cde}_{kln}(\vk_3,\vk_4,\vk_1+\vk_2) \\
&{}\qquad + \gamma^{ace}_{ikm}(\vk_1,\vk_3,-\vk_1-\vk_3) \, t_{mn}(\vk_1+\vk_3) \, \gamma^{bde}_{jln}(\vk_2,\vk_4,\vk_1+\vk_3) \\
&{}\qquad\qquad + \gamma^{ade}_{ilm}(\vk_1,\vk_4,-\vk_1-\vk_4) \, t_{mn}(\vk_1+\vk_4) \gamma^{bce}_{jkn}(\vk_2,\vk_3,\vk_1+\vk_4)
\biggr\} \\
-2 g^2 & \biggl\{
f^{abe} f^{cde} \delta_{ij} \delta_{kl}
\bigl[\Omega(\vk_1) - \Omega(\vk_2)\bigr] F(\vk_1+\vk_2) \bigl[\Omega(\vk_3) - \Omega(\vk_4)\bigr] \\
&{}\qquad + f^{ace} f^{bde} \delta_{ik} \delta_{jl}
\bigl[\Omega(\vk_1) - \Omega(\vk_3)\bigr] F(\vk_1+\vk_3) \bigl[\Omega(\vk_2) - \Omega(\vk_4)\bigr] \\
&{}\qquad\qquad + f^{ade} f^{bce} \delta_{il} \delta_{jk}
\bigl[\Omega(\vk_1) - \Omega(\vk_4)\bigr] F(\vk_1+\vk_4) \bigl[\Omega(\vk_2) - \Omega(\vk_3)\bigr]
\biggr\} ,
\end{split}
\ee
where $\Omega(\vk)$ is the gluon energy defined by the static gluon propagator
\[
\vev{A_i^a(\vk) A_j^b(\vq)} = \delta^{ab} \, \frac{t_{ij}(\vk)}{2\Omega(\vk)} \, \deltabar(\vp+\vq),
\]
with
\[
t_{ij}(\vk) = \delta_{ij} - \frac{k_i k_j}{\vk^2}
\]
being the transverse projector. In the equation for the four-gluon variational kernel
the Coulomb interaction kernel $F(\vk)$ appears, which is given in Eq.~(\ref{eq:F}) below. Furthermore
\be\label{t3}
T^{abc}_{ijk}(\vp,\vq,\vk) = \I* \, f^{abc}\bigl[ \delta_{ij} (p-q)_k + \delta_{jk} (q-k)_i + \delta_{ki} (k-p)_j \bigr]
\ee
and
\be\label{t4}
T^{abcd}_{ijkl} =
f^{abe} f^{cde} (\delta_{ik} \, \delta_{jl} - \delta_{il} \, \delta_{jk}
+ f^{ace} f^{bde} (\delta_{ij} \, \delta_{kl} - \delta_{jk} \, \delta_{il})
+ f^{ade} f^{bce} (\delta_{ij} \, \delta_{kl} - \delta_{ik} \, \delta_{jl})
\ee
denote the tensor structures of the three- and
four-gluon couplings occurring in the Yang--Mills Hamiltonian.
The four-gluon kernel \Eqref{k4} is schematically illustrated in Fig.~\ref{fig:4gk}.
The variational equation for the two-gluon kernel $\omega$ can be combined with the
CRDSE for the gluon propagator $\Omega$, resulting in the so-called gap equation \cite{Feuchter:2004mk,Campagnari:2010wc}.

Lattice data for the gluon propagator \cite{Burgio:2008jr} can be well fitted by Gribov's formula
\be\label{input1}
\Omega(\vk) = \sqrt{\vk^2 + \frac{m_A^4}{\vk^2}}
\ee
with an effective mass $m_A\simeq880$\,MeV (for $\Nc=2$). Alternatively, this can be
expressed via the so-called Coulomb string tension as $m_A^2=0.6\sigma_\mathrm{C}$ [see \Eqref{eq:F} below].

\sffamily
A comment is here in order: The lattice data do not really go into the deep IR. So the precise value
of the IR exponent $\alpha$ of $\Omega(p\to0)\sim p^{-\alpha}$ cannot be accurately determined
from the lattice data of Ref.~\cite{Burgio:2008jr}. However, the same IR exponent is also found in the
continuum calculation \cite{Epple:2006hv}, see below.
\rmfamily

The ghost propagator \Eqref{ghp-def} is represented in momentum space as
\[
G(\vk) = \vev{G_A} = \frac{d(\vk)}{g \, \vk^2} \, ,
\]
where $d(\vk)$ is the ghost form factor.
Assuming the so-called horizon condition $d^{-1}(0)=0$ and a bare ghost-gluon vertex
one finds from the variational calculation
\sffamily
carried out with a Gaussian vacuum wave functional \cite{Feuchter:2004mk} two scaling-type
solutions:\footnote{Note that this is different from Landau gauge, where one finds a
`scaling' and a `decoupling' solution but only the latter is consisten with lattice data.}
one with a gluon IR exponent $\alpha=0.6$ \cite{Feuchter:2004mk} and one with $\alpha=1$ \cite{Epple:2006hv}.
Both solutions are also obtained in an IR analysis of the equations of motion (gap equation
and ghost DSE) \cite{Schleifenbaum:2006bq}. We prefer here to use the solution with
$\alpha=1$ as input for the CRDSEs since this solution not only seems to be in better
agreement with the lattice data for the gluon propagator but leads
\rmfamily
also to a linearly rising non-Abelian Coulomb potential
\begin{align}\label{eq:F}
F(\vp) = \vp^2 G^2(\vp) \xrightarrow{\vp\to0} \frac{8\pi\sigmacoul}{\vp^4} ,
\end{align}
which again is consistent with the lattice data. Here $\sigmacoul$ is the Coulomb string tension,
which is found on the lattice to be about two to three times larger than the Wilson string tension.
For later use we also note that the ghost form factor $d$ obtained in Ref.~\cite{Epple:2006hv}
for the $\alpha=1$ solution can be fitted by \cite{Campagnari:2010wc}
\be\label{input2}
d(x) = a \sqrt{\frac{1}{x^2}+\frac{1}{\ln(x^2+c^2)}} \, , \qquad x^2\equiv\frac{\vp^2}{\sigmacoul}, \qquad c\simeq 4, \qquad a\simeq5.
\ee
To simplify the numerical solution of the CRDSEs we will parameterize the gluon energy by
the Gribov formula Eq.~\eqref{input1} and the ghost form factor by Eq.~\eqref{input2}.

Equations~(\ref{input1}) and (\ref{input2}) constitute the input of our calculations.
They also set the scale and we represent all results in units of the Coulomb string tension $\sigmacoul$.
All calculations were done for $\mathrm{SU}(2)$. The coupling $g$ was set to $3.5$.
This corresponds to a renormalization point of $\mu=2.4\sqrt{\sigmacoul}$~\cite{Epple:2006hv}.


\section{Canonical Recursive Dyson--Schwinger Equations for Vertex Functions}

The CRDSEs for the vertices have been derived in Ref.~\cite{Campagnari:2010wc},
to which we refer the reader for the details; here we give merely a short summary of the
derivation and quote the relevant one-loop results.

\subsection{Ghost-Gluon Vertex}

The Faddeev--Popov operator \Eqref{ggvdse0} can be inverted to give
the operator identity
\be\label{ggv1}
G_A(1,2) = G_0(1,2) - G_A(1,3) A(4) \widetilde{\Gamma}_0(3,5;4) G_0(5,2) .
\ee
In \Eqref{ggv1}, $\widetilde\Gamma_0$
is the bare ghost-gluon vertex [see \Eqref{ggv0} below], and $G_0=G_{A=0}$ is the bare ghost propagator.

\begin{figure}
\centering
\includegraphics[width=.5\linewidth]{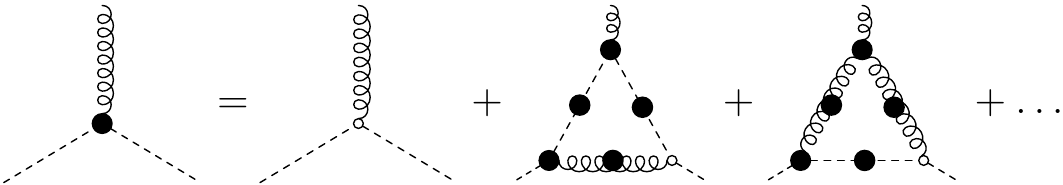}
\caption{\label{fig:dse_ggv_1}The CRDSE \protect\eqref{ggvdse1} for the ghost-gluon vertex, arising
from the operator identity Eq.~\protect\eqref{ggv1}. Wiggly and dashed lines represent the bare
gluon and ghost propagators, respectively. If these lines are augmented by a full dot they represent full propagators.
Empty and full (fat) dots stand for bare and full (dressed, one-particle irreducible) vertices.}
\end{figure}

\begin{figure}
\centering
\includegraphics[width=.5\linewidth]{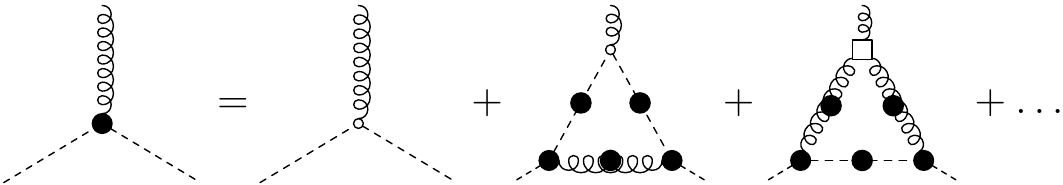}
\caption{\label{fig:dse_ggv_2}Alternative form of the CRDSE for the ghost-gluon vertex, arising
from the functional identity Eq.~\protect\eqref{mad3}. For notation see caption of Fig.~\protect\ref{fig:dse_ggv_1}.
Furthermore, empty boxes represent variational kernels.}
\end{figure}

Multiplying \Eqref{ggv1} by the (spatial) gauge field $A$ and taking the expectation value
yields for the ghost-gluon vertex \Eqref{ggv-def} at one-loop level the following CRDSE
\be\label{ggvdse1}
\begin{split}
\widetilde{\Gamma}(1,2;3) =
\widetilde{\Gamma}_0(1,2;3) &+ \widetilde{\Gamma}(1,4;6') G(4,4') \widetilde{\Gamma}(4',5;3) G(5,5') \widetilde{\Gamma}_0(5',2;6) D(6,6') \\
&+ \widetilde{\Gamma}(1,6;4) D(4,4') \Gamma(4',5;3) D(5,5') \widetilde{\Gamma}_0(6',2;5') G(6,6') + \dots,
\end{split}
\ee
which is represented diagrammatically in Fig.~\ref{fig:dse_ggv_1}.
An alternative equation can be obtained by putting $K=G_A$ in \Eqref{mad3}: this leads
to the CRDSE
\be\label{ggvdse2}
\begin{split}
\widetilde{\Gamma}(1,2;3) =
\widetilde{\Gamma}_0(1,2;3) &+ \widetilde{\Gamma}(1,4;6') G(4,4') \widetilde{\Gamma}_0(4',5;3) G(5,5') \widetilde{\Gamma}(5',2;6) D(6,6') \\
&+ \widetilde{\Gamma}(1,6;4) D(4,4') \gamma(4',5,3) D(5,5') \widetilde{\Gamma}(6',2;5') G(6,6') + \dots,
\end{split}
\ee
represented diagrammatically in Fig.~\ref{fig:dse_ggv_2}.
The two CRDSEs for the ghost-gluon vertex read schematically (trivial color factor $f^{abc}$ suppressed)
\[
\widetilde{\Gamma}_{i}(\vp,\vq;\vk) = \widetilde{\Gamma}_{0,i}(\vp,\vq;\vk)
  +\Sigma^{\text{Ab}}_{i}(\vp,\vq;\vk)
  +\Sigma^{\text{non-Ab}}_{i}(\vp,\vq;\vk)
  + \ldots
\]
where the bare vertex $\widetilde{\Gamma}_{0,i}$ is given by
\be\label{ggv0}
\widetilde{\Gamma}_{0,i}(\vp,\vq;\vk) = \I * g \, t_{ij}(\vk) p_j .
\ee
Furthermore, $\Sigma^{\text{Ab}}_i$ and $\Sigma^{\text{non-Ab}}_i$ represent
the second and third diagrams, respectively, on the r.h.s.~of the CRDSEs shown in Figs.~\ref{fig:dse_ggv_1}
and \ref{fig:dse_ggv_2}. For each version of the two CRDSEs the ellipses denote
different diagrams neglected in our truncation, namely all two-loop diagrams (which only
appear in the CRDSE with a three-gluon kernel) and diagrams with non-primitively
divergent Green functions.
At one-loop level these two equations differ by the leg attached to the bare vertex: the
anti-ghost in Fig.~\ref{fig:dse_ggv_1} and the gluon in Fig.~\ref{fig:dse_ggv_2}.
Furthermore, the full (dressed) three-gluon vertex of the second loop diagram in Fig.~\ref{fig:dse_ggv_1}
is replaced in Fig.~\ref{fig:dse_ggv_2} by the variational kernel $\gamma_3$. As we will
see in the next subsection, at leading order the dressed three-gluon vertex is given by the variational kernel
$\gamma_3$, see Fig.~\ref{fig:dse_3gv} or \Eqref{dse11}. In the numerical calculation we will solve
the CRDSE for the ghost-gluon vertex given by \Eqref{ggvdse1} (Fig.~\ref{fig:dse_ggv_1})
but replace the three-gluon vertex by the variational kernel $\gamma_3$. The resulting
CRDSE differs then from the one shown in Fig.~\ref{fig:dse_ggv_2} only by the leg attached
to the bare vertex.

Due to the transversality of the gluon propagator, the colour and Lorentz structure of the full
ghost-gluon vertex is the same as the bare one \Eqref{ggv0}. Hence there is only one
relevant dressing function for the full ghost-gluon vertex, which can be chosen as
\[
\widetilde{\Gamma}_{i}(\vp,\vq;\vk) = \I* g t_{ij}(\vk) p_j D^{\bar{c}cA}(\vp,\vq;\vk) .
\]
The arguments of the dressing function $D^{\bar{c}cA}(\vp,\vq;\vk)$ are the incoming three-momenta
of the anti-ghost, the ghost and the gluon legs. Alternatively
also the moduli of the anti-ghost and gluon momenta and the angle between them will be used:
$D^{\bar{c}{c}A}(\abs{\vp},\abs{\vk},\alpha)$. To obtain a scalar integral equation
for the dressing function, the CRDSE~\eqref{ggvdse1} [or \Eqref{ggvdse2}] is contracted
with the projector 
\be\label{eq:projector-ggv}
 P^{\bar{c}{c}A}_{i}:=-\frac{\I}{g}\,\frac{p_i}{p_{j} t_{jl}(\vk) p_{l}} \, .
\ee
This results in the following integral equation:
\begin{align}\label{eq:ghgDSE}
 D^{\bar{c}cA}(\vp,\vq;\vk)=1+\Sigma^{\text{Ab}}(\vp,\vq;\vk)+\Sigma^{\text{non-Ab}}(\vp,\vq;\vk),
\end{align}
where the two contributions $\Sigma^{\text{Ab}}$ and $\Sigma^{\text{non-Ab}}$ (without Lorentz index) correspond to the projected diagrams $\Sigma^{\text{Ab}}_i$ and $\Sigma^{\text{non-Ab}}_i$, respectively. The explicit expressions for the kernels are given in the Appendix.
Although the projector \Eqref{eq:projector-ggv} is ill-defined for $\vp=\pm\vk$,
the projected diagrams are free of kinematical singularities.


\subsection{Three-Gluon Vertex}\label{sec:3g-DSE}

\begin{figure}
\centering
\includegraphics[width=.9\linewidth]{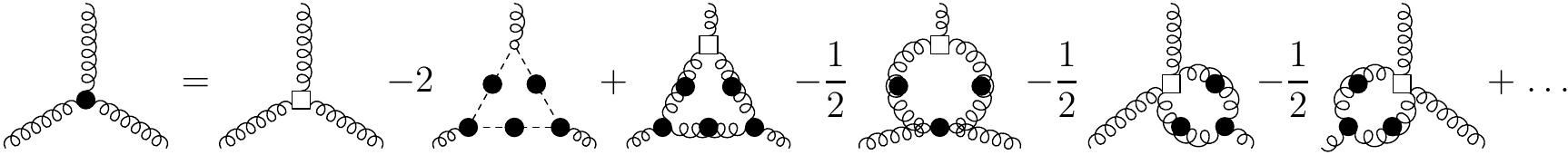}
\caption{\label{fig:dse_3gv}CRDSE for the three-gluon vertex.}
\end{figure}

The CRDSE for the three-gluon vertex $\Gamma_3$ [\Eqref{3gv-def}] is obtained from \Eqref{mad3} by taking $K[A]$
to be the product of two gauge fields. It reads \cite{Campagnari:2010wc}
\begin{align}\displaybreak[1]
\Gamma(1,2,3) ={}& \gamma(1,2,3) - 2 \widetilde{\Gamma}_0(1;4,5) \, G(4',4) \, G(5,5') \, G(6',6)
\widetilde{\Gamma}(2;6,4') \, \widetilde{\Gamma}(3;5',6') \nn \\
& + \gamma(1,4,5) \, D(4,4') \, D(5,5') \, D(6,6') \, \Gamma(2,4',6) \, \Gamma(3,5',6') \nn \\
&-\frac12 \: \gamma(1,4,5) \, D(4,4') \, D(5,5') \, \Gamma(4',5',2,3) \nn \\
&- \frac12 \bigl[ \gamma(1,2,4,5) \, D(4,4') \, D(5,5') \, \Gamma(4',5',3) + 2 \leftrightarrow 3 \bigr] + \dots\label{dse11},
\end{align}
and is represented in Fig.~\ref{fig:dse_3gv}.
In Ref.~\cite{Campagnari:2010wc} this equation has been studied at leading infrared (IR) order, i.e.~only the ghost triangle was considered. In this work we will
consider also the gluonic contributions given by the gluon triangle and the three swordfish diagrams. The full four-gluon
vertex will be replaced by the variational kernel, and the r.h.s.~of Fig.~\ref{fig:dse_3gv} will
be properly Bose-symmetrized in the three gluon legs.
Writing out explicitly the Lorentz indices and the momentum variables but suppressing
a trivial color factor $f^{abc}$ the CRDSE of the three-gluon vertex reads schematically
\be\label{dse11a}
\begin{split}
 \Gamma_{ijk}(\vp,\vq,\vk) ={}& \gamma_{ijk}(\vp,\vq,\vk)
  -2\Sigma^{\text{gh-tr}}_{ijk}(\vp,\vq,\vk)
  +\Sigma^{\text{gl-tr}}_{ijk}(\vp,\vq,\vk) \\
  &-\frac{1}{2}\Sigma^{\text{sw}_1}_{ijk}(\vp,\vq,\vk)
  -\frac{1}{2}\Sigma^{\text{sw}_2}_{ijk}(\vp,\vq,\vk)
  -\frac{1}{2}\Sigma^{\text{sw}_3}_{ijk}(\vp,\vq,\vk)
  + \ldots
\end{split}
\ee
where each term represents a diagram of Fig.~\ref{fig:dse_3gv}.
The variational kernel $\gamma_3$ is given by Eqs.~\eqref{k3} and \eqref{t3}.
The ellipses represent diagrams neglected within the present truncation, namely two-loop
terms and a diagram containing the ghost-gluon four-point function. For the full three-gluon
vertex we will assume the same Lorentz structure as for the bare one
\be\label{v3proj}
 \Gamma^{abc}_{ijk}(\vp,\vq,\vk)=g\,D^{A^3}(\vp,\vq,\vk)T^{abc}_{ijk}(\vp,\vq,\vk).
\ee
Other dressing functions do exist, but it was shown in the case of the Landau gauge by
direct calculation \cite{Eichmann:2014xya} and by comparison with lattice results \cite{Blum:2014gna} that they are very small. This motivates the use of the same approximation here.
The arguments of the dressing function $D^{A^3}(\vp,\vq,\vk)$ are the incoming three-momenta.
Also here the moduli of the first two momenta and the angle between them will be used as well: $D^{A^3}(|\vp|,|\vq|,\alpha)$.
To obtain a scalar integral equation for the dressing function we contract the CRDSE~\eqref{dse11}
with the following projector\footnote{%
Note that we project the full vertex $\Gamma_3$ onto the kernel $\gamma_3$ \Eqref{k3};
we do \emph{not} project it onto the perturbative vertex, which is given by \Eqref{k3}
with $\Omega(\vp)$ replaced by $\abs{\vp}$.}
\begin{align}\label{eq:projector-3gv}
P^{A^3,abc}_{ijk}(\vp,\vq,\vk) \coloneq
\frac{\gamma^{abc}_{lmn}(\vp,\vq,\vk)t_{li}(\vp)t_{mj}(\vq)t_{nk}(\vk)}{\gamma^{def}_{opq}(\vp,\vq,\vk)t_{oo'}(\vp)t_{pp'}(\vq)t_{qq'}(\vk)\gamma^{def}_{o'p'q'}(\vp,\vq,\vk)}.
\end{align}
On the left-hand side of the CRDSE~\eqref{dse11a} we get then
\[
 P^{A^3,abc}_{ijk}(\vp,\vq,\vk)\Gamma^{A^3,abc}_{ijk}(\vp,\vq,\vk)=D^{A^3}(\vp,\vq,\vk)\frac{\Omega(\vp)+\Omega(\vq)+\Omega(\vk)}{2}.
\]
On the right-hand side the term from the variational kernel becomes just $1$ (i.e.~a momentum
independent constant). This is important to handle the divergences on the right-hand side,
because now a simple momentum subtraction can be used. (For other projections the divergent
integrals have prefactors that depend on the external momenta and momentum subtraction does not work.)
The subtraction point is chosen as $|\vp|=|\vq|=|\vk|=|\vpn|=p_0$ with $|\vpn|$ in the UV:
\begin{align}\label{eq:selfenergy_sub_proj}
 &D^{A^3}(\vp,\vq,\vk)\frac{\Omega(\vp)+\Omega(\vq)+\Omega(\vk)}{2}-D^{A^3}(p_0,p_0,2\pi/3)\frac{3\Omega(p_0)}{2}=\nnnl
 &\quad\Sigma(\vp,\vq,\vk) \frac{\Omega(\vp)+\Omega(\vq)+\Omega(\vk)}{2}-\Sigma(p_0,p_0,p_0)\frac{3\Omega(p_0)}{2}=\Sigma^{\text{sub,proj}}(\vp,\vq,\vk),
\end{align}
where $\Sigma(\vp,\vq,\vk)$ denotes the sum of all projected integrals with the $\gamma$-dependent part from the projection factored out.
The renormalization condition is chosen as $D^{A^3}(p_0,p_0,2\pi/3)=\gamma_3(p_0,p_0,2\pi/3)=2/3\Omega(p_0)$.
The result for the three-gluon vertex dressing reads then:
\begin{align}\label{eq:DAAA}
 D^{A^3}(\vp,\vq,\vk)=\frac{2}{\Omega(\vp)+\Omega(\vq)+\Omega(\vk)}\left(1+\Sigma^{\text{sub,proj}}(\vp,\vq,\vk)\right),
\end{align}
where $\Sigma^{\text{sub,proj}}$ is given by \Eqref{eq:selfenergy_sub_proj}. In our numerical calculations we used $p_0=600 \sqrt{\smash[b]{\sigmacoul}}$.

The full three-gluon vertex is totally symmetric with respect to a permutation of the
external gluon legs. The r.h.s.~of the CRDSE~\eqref{dse11} and the corresponding diagrams
in Fig.~\ref{fig:dse_3gv} do not respect this symmetry due to the truncation. We restore this symmetry
by averaging the final integral equation over inequivalent permutations of the external
gluon legs, resulting in
\be\label{3gvdsen}
\begin{split}
 D^{A^3,\mathrm{symm}}&(\vp,\vq,\vk)=\frac{1}{3}\left(D^{A^3}(\vp,\vq,\vk)+D^{A^3}(\vk,\vp,\vq)+D^{A^3}(\vq,\vk,\vp)\right)\\
 ={}&\frac{2}{\Omega(\vp)+\Omega(\vq)+\Omega(\vk)} \\
 \times\biggl\{ 1 &-\frac{2}{3}\Bigl(\Sigma^{\text{gh-tr,sub}}(\vp,\vq,\vk)+\Sigma^{\text{gh-tr,sub}}(\vk,\vp,\vq)+\Sigma^{\text{gh-tr,sub}}(\vq,\vk,\vp)\Bigr)\\
  &+\frac{1}{3}\Bigl(\Sigma^{\text{gl-tr,sub}}(\vp,\vq,\vk)+\Sigma^{\text{gl-tr,sub}}(\vk,\vp,\vq)+\Sigma^{\text{gl-tr,sub}}(\vq,\vk,\vp)\Bigr)\\
  &-\frac{1}{6}\Bigl(\Sigma^{\text{sw}_1\text{,sub}}(\vp,\vq,\vk)+\Sigma^{\text{sw}_1\text{,sub}}(\vk,\vp,\vq)+\Sigma^{\text{sw}_1\text{,sub}}(\vq,\vk,\vp)\Bigr)\\
  &-\frac{1}{3}\Bigl(\Sigma^{\text{sw}_2\text{,sub}}(\vp,\vq,\vk)+\Sigma^{\text{sw}_2\text{,sub}}(\vk,\vp,\vq)+\Sigma^{\text{sw}_2\text{,sub}}(\vq,\vk,\vp)\Bigr) \biggr\}
  + \ldots
\end{split}
\ee
Due to this symmetrization the two swordfish diagrams with variational four-gluon vertices kernels can
be subsumed (diagrams five and six in \fref{fig:dse_3gv}).

To alleviate the algebraic manipulations performed before creating the kernel files for the numeric code, the expression \Eqref{k4} is split into three parts:
\begin{align}
 \gamma^{(1),abcd}_{ijkl}(&\vk_1,\vk_2,\vk_3,\vk_4) = \frac{2 \, g^2 \, T^{abcd}_{ijkl}}{\bigl[ \Omega(\vk_1) + \Omega(\vk_2) + \Omega(\vk_3) + \Omega(\vk_4) \bigr]} \\
\gamma^{(2),abcd}_{ijkl}(&\vk_1,\vk_2,\vk_3,\vk_4) =-\frac12 \frac{1}{\bigl[ \Omega(\vk_1) + \Omega(\vk_2) + \Omega(\vk_3) + \Omega(\vk_4) \bigr]}\nnnl
 &{}\qquad\times\biggl\{\gamma^{abe}_{ijm}(\vk_1,\vk_2,-\vk_1-\vk_2) \, t_{mn}(\vk_1+\vk_2) \, \gamma^{cde}_{kln}(\vk_3,\vk_4,\vk_1+\vk_2) \nnnl
&{}\qquad\qquad + \gamma^{ace}_{ikm}(\vk_1,\vk_3,-\vk_1-\vk_3) \, t_{mn}(\vk_1+\vk_3) \, \gamma^{bde}_{jln}(\vk_2,\vk_4,\vk_1+\vk_3) \nnnl
&{}\qquad\qquad + \gamma^{ade}_{ilm}(\vk_1,\vk_4,-\vk_1-\vk_4) \, t_{mn}(\vk_1+\vk_4) \gamma^{bce}_{jkn}(\vk_2,\vk_3,\vk_1+\vk_4)
\biggr\} \\
\gamma^{(3),abcd}_{ijkl}(&\vk_1,\vk_2,\vk_3,\vk_4) = \frac{-2 g^2}{\bigl[ \Omega(\vk_1) + \Omega(\vk_2) + \Omega(\vk_3) + \Omega(\vk_4) \bigr]}\nnnl
&{}\qquad\times\biggl\{
f^{abe} f^{cde} \delta_{ij} \delta_{kl}
\bigl[\Omega(\vk_1) - \Omega(\vk_2)\bigr] F(\vk_1+\vk_2) \bigl[\Omega(\vk_3) - \Omega(\vk_4)\bigr] \nnnl
&{}\qquad\qquad + f^{ace} f^{bde} \delta_{ik} \delta_{jl}
\bigl[\Omega(\vk_1) - \Omega(\vk_3)\bigr] F(\vk_1+\vk_3) \bigl[\Omega(\vk_2) - \Omega(\vk_4)\bigr] \nnnl
&{}\qquad\qquad + f^{ade} f^{bce} \delta_{il} \delta_{jk}
\bigl[\Omega(\vk_1) - \Omega(\vk_4)\bigr] F(\vk_1+\vk_4) \bigl[\Omega(\vk_2) - \Omega(\vk_3)\bigr]
\biggr\}.
\label{4gk-3}
\end{align}
From the IR behaviour of the gluon energy $\Omega$ [\Eqref{input1}] and of the Coulomb kernel $F$ [\Eqref{input2}]
follows that the third part $\gamma_4^{(3)}$ [\Eqref{4gk-3}] of the four-gluon kernel
behaves quantitatively like $p^{-5}$ for $p\to0$.
This is the same degree of IR divergence as expected from the analysis of the ghost box of
the four-gluon vertex DSE \cite{Huber:2007kc}. As a consequence, the swordfish diagrams
containing both the variational four-gluon kernel and one full three-gluon vertex
[last two terms on the r.h.s.~of \Eqref{dse11}/\fref{fig:dse_3gv}] diverge like
$p^{-5+2\times1+3-3}=p^{-3}$ and contribute at the same order as the ghost triangle
[the second term on the r.h.s.~of \Eqref{dse11}/\fref{fig:dse_3gv}]. This comes somewhat
unexpected as typically ghost dominance is manifest.

In Ref.~\cite{Campagnari:2010wc}
the three-gluon vertex was calculated in the symmetric momentum
configuration, for which $\vk_i^2=p^2$ and $\vk_i\cdot\vk_j=-p^2/3$, $i\neq j$, and
$\gamma^{(3),abcd}_{ijkl}(\vk_1,\vk_2,\vk_3,\vk_4)=0$ holds.
This considerably simplifies the variational four-gluon kernel. Here, however, we will resolve the
full momentum dependence of the three-gluon vertex.
Due to the quite involved expression for the variational four-gluon kernel the derivation of the final
integral kernels becomes very cumbersome and more complicated than in the Landau gauge. Thus the use
of a computer algebra system is almost unavoidable and we used the \textit{Mathematica} \cite{Wolfram:2004} package \textit{DoFun} for this task \cite{Alkofer:2008nt,Huber:2011qr}.


\section{Numerical Results}

For the numerical calculation the vertices are put on a grid. Up to 40 points for each
of the external momenta and up to 18 points for the external angle were used. For
intermediate points linear interpolation was employed. Naturally also values outside of
the grid are required.\footnote{For the angle this does not apply as it is a bounded variable.
However, to avoid finite $0/0$ expressions, which are difficult to handle numerically,
the lowest and highest values are slightly higher and lower, respectively.}
In the IR the boundary values were used. For the ghost-gluon vertex this is a trivial
choice, as it becomes constant in the IR. Also for the three-gluon vertex this prescription
was adopted. The only diagram that could be affected by this choice is the swordfish diagram
with a full three-gluon vertex.\footnote{The other diagram with a full three-gluon vertex
is the gluon-triangle. It is IR suppressed and thus any effect on its IR behavior does
not couple back on the vertex itself.} However, we demonstrate in \fref{fig:3g-compIR}
that for the chosen IR parameters it is not affected either and follows the expected
power law. In the UV the boundary value is taken as well. This choice does not respect
the anomalous dimensions of the vertices. To clarify its influence we varied the grid
size for the ghost-gluon vertex but found no visible change, which validates this procedure a posteriori.

The three-dimensional integrals are done using spherical coordinates:
\begin{align}
 \int d^3 \omega = \int d\omega\, \omega^2 \int d\theta_2  \sin(\theta_2) \int d\theta_1.
\end{align}
To avoid possible problems with the denominators of the integrands, we split the integration
regions at their zeros \cite{Schleifenbaum:2004dt}. Thus the radial integration contains
three regions with up to 70 points each and the angle integrations two regions with up
to 35 points. Besides this aspect the integration is rather trivial and a simple Gauss--Legendre
quadrature is sufficient. The IR/UV cutoff was set a factor 100/10 times lower/higher
than the lowest/highest grid point. The cutoff independence was verified by varying the
cutoff by a factor of 10, which has no effect.

As starting values we used for the ghost-gluon vertex the bare vertex and for the three-gluon
vertex the ghost-triangle-only calculation. The final result is obtained by a fixed point iteration.
All calculations were performed with the \textit{CrasyDSE} framework~\cite{Huber:2011xc}.
Further numerical details can also be found in Ref. \cite{Huber:2011xc} and references therein.

\subsection{\label{sec:ghg_res}Ghost-Gluon Vertex}

\begin{figure}[tb]
 \begin{center}
 \includegraphics[width=0.48\textwidth]{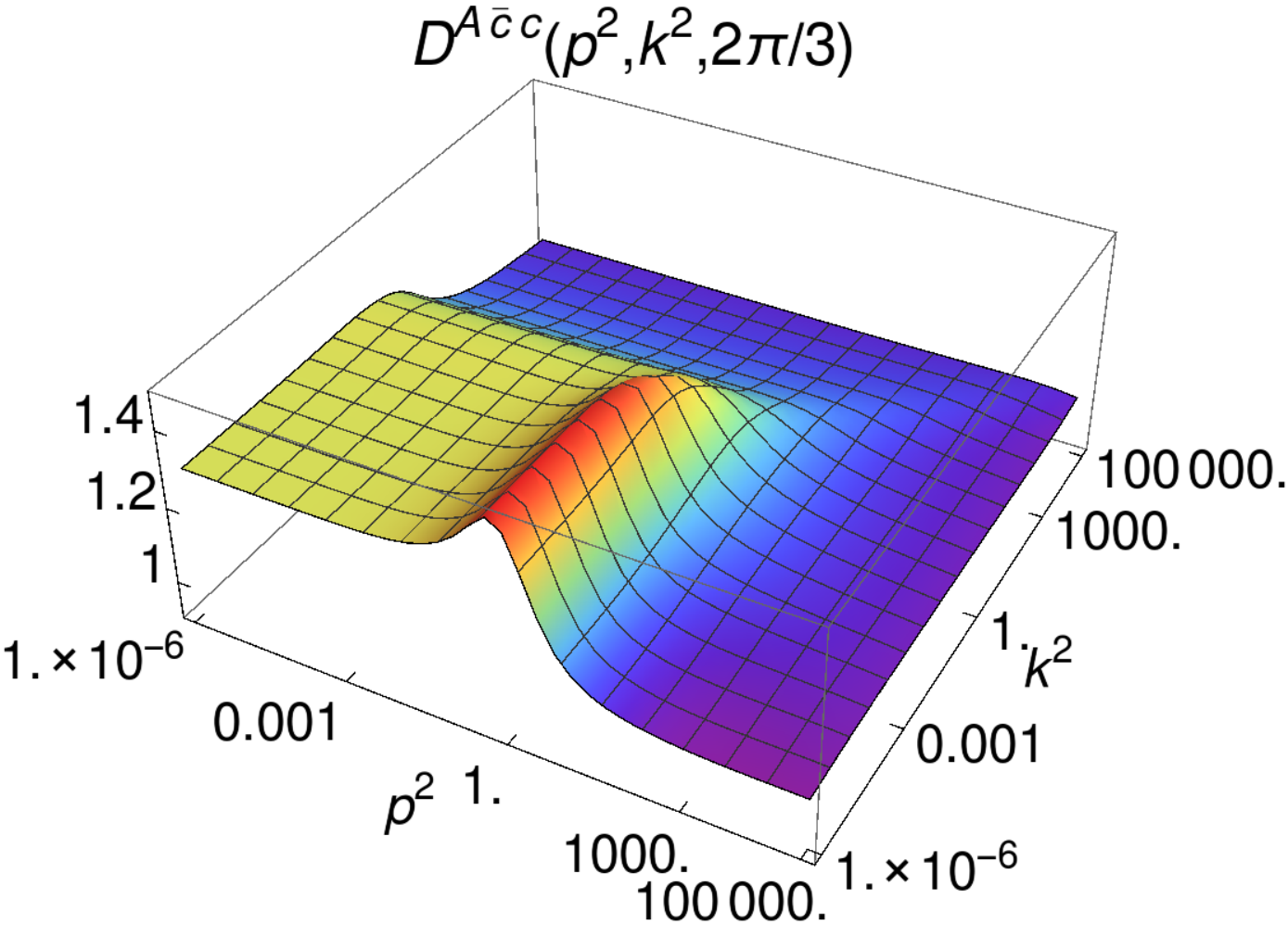} 
 \includegraphics[width=0.48\textwidth]{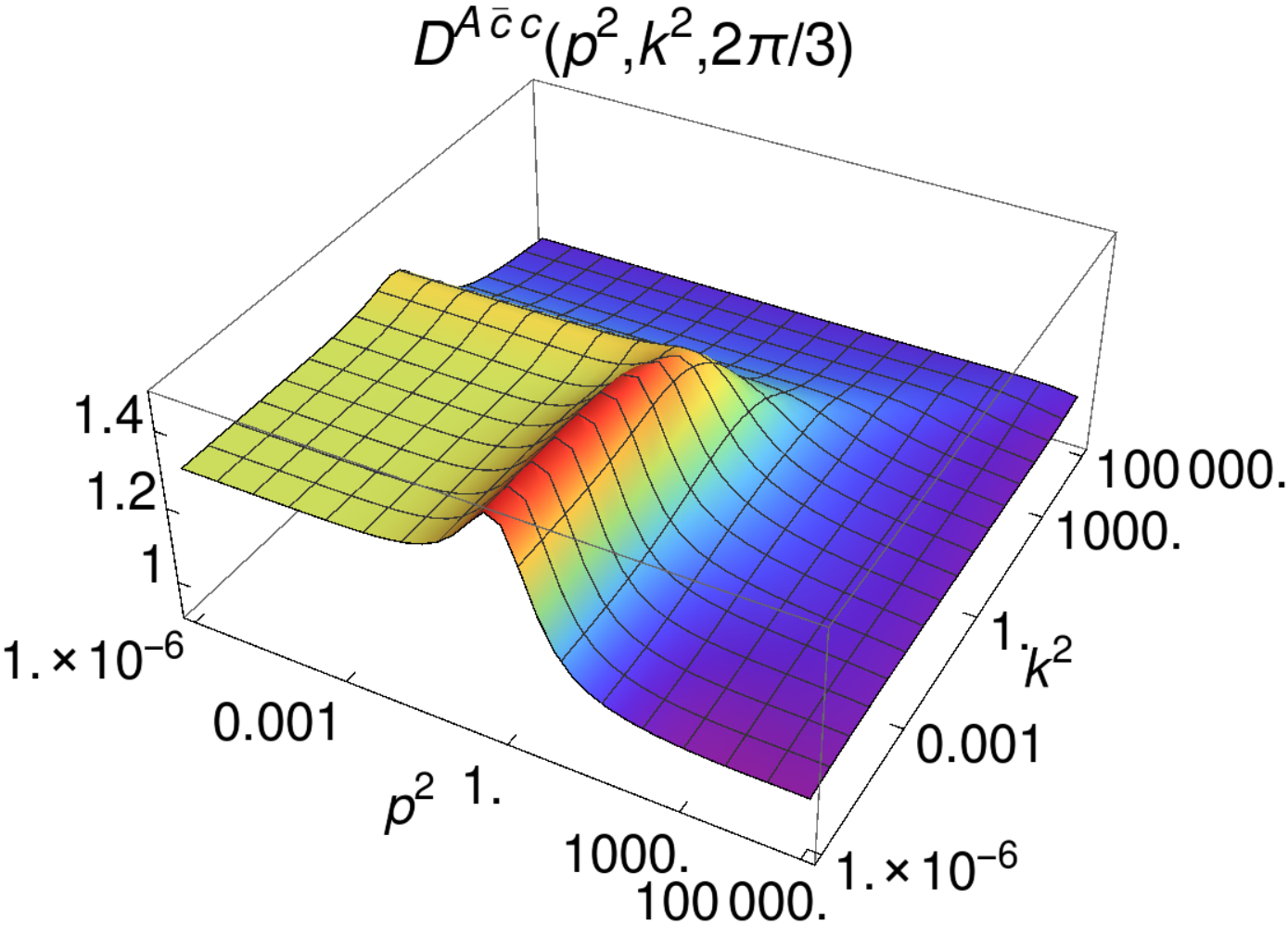}
 \caption{\label{fig:ghg-3d}Dressing function of the ghost-gluon vertex. The anti-ghost momentum is denoted by $p$, the gluon momentum by $k$. \textit{Left}: Anti-ghost legs attached to bare vertices, see Fig.~\ref{fig:dse_ggv_1}. \textit{Right}: Gluon legs attached to bare vertices, see Fig.~\ref{fig:dse_ggv_2}.}
 \end{center}
\end{figure}

\begin{figure}[tb]
 \begin{center}
 \includegraphics[width=0.48\textwidth]{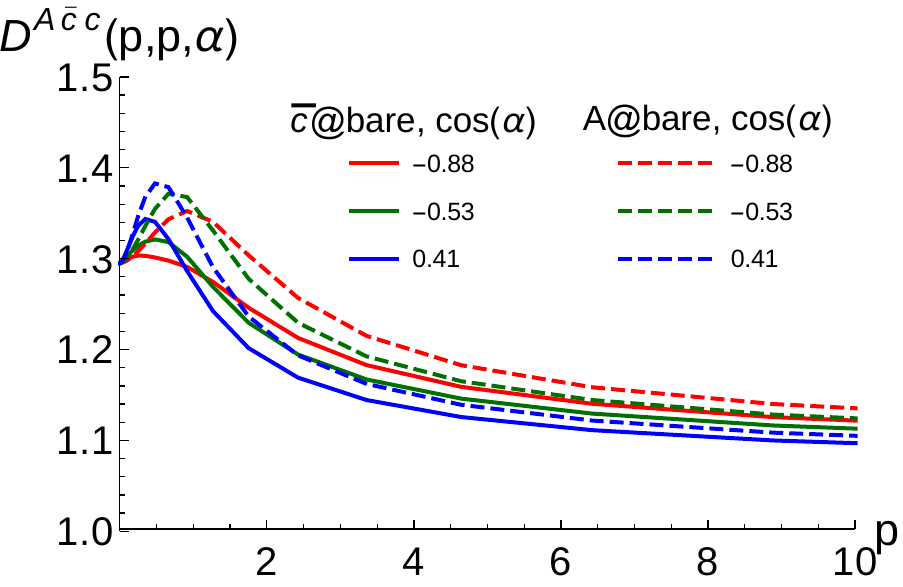} 
 \includegraphics[width=0.48\textwidth]{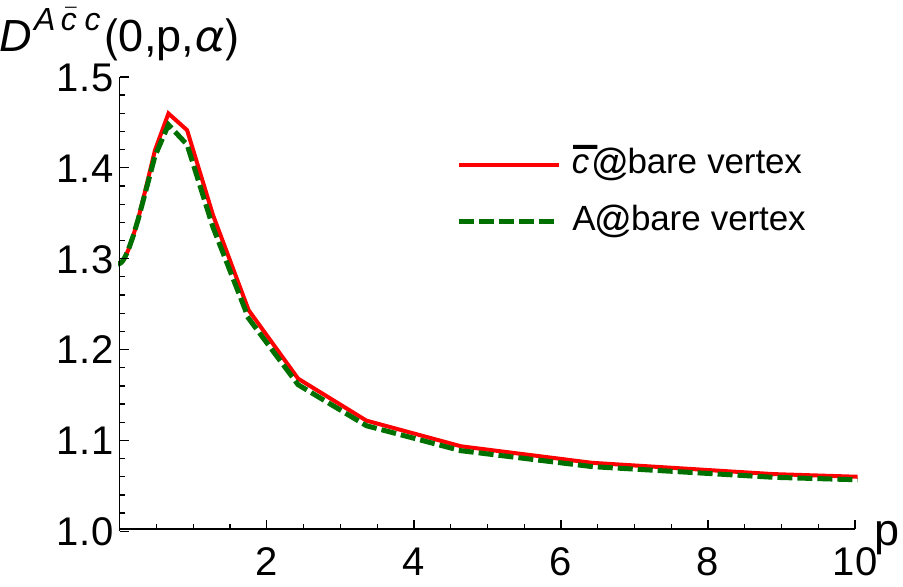}
 \caption{\label{fig:ghg_comp_cbA}Comparison of the results from the two different ghost-gluon vertex CRDSEs. Continuous/Dashed lines are from the versions with the anti-ghost/gluon legs attached to the bare vertices. \textit{Left}: Equal anti-ghost and gluon momenta, different angles. \textit{Right}: Zero gluon momentum.}
 \end{center}
\end{figure}

Figure \ref{fig:ghg-3d} shows the dressing function of the ghost-gluon vertex
as function of the modulus of the two external momenta for a fixed angle of roughly $2\pi/3$.
The two panels show the dressing function of the ghost-gluon vertex obtained from the CRDSE~\eqref{ggvdse1}
[Fig.~\ref{fig:dse_ggv_1}] and \Eqref{ggvdse2} [Fig.~\ref{fig:dse_ggv_2}], respectively.
In general the differences are small and largest for the ridge with constant gluon momentum.
A detailed comparison of the results from the two different ghost-gluon vertex CRDSEs is
shown in \fref{fig:ghg_comp_cbA} for specific momentum configurations.

\begin{figure}[tb]
 \begin{center}
 \includegraphics[width=0.48\textwidth]{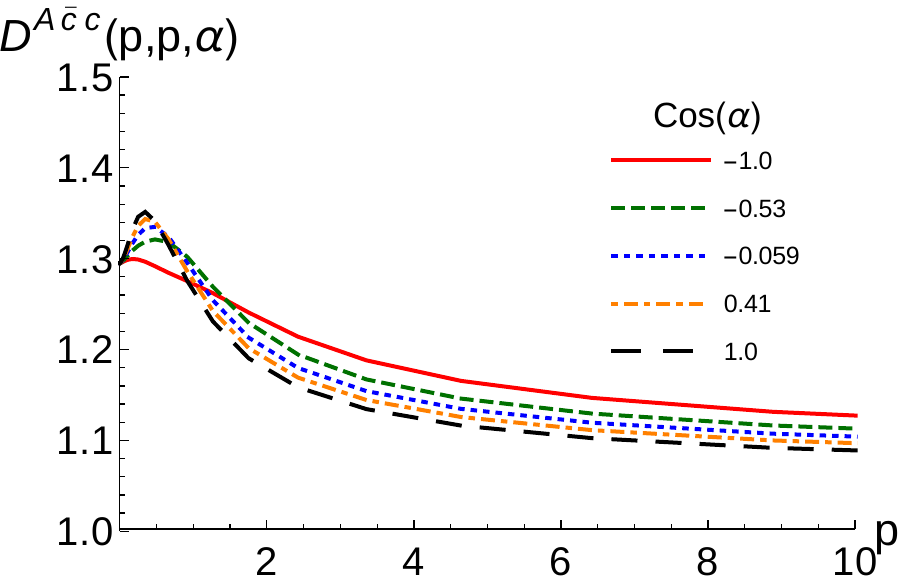}
 \caption{\label{fig:ghg_equalMoms}Ghost-gluon vertex dressing function for equal momenta and various angles.}
 \end{center}
\end{figure}

\begin{figure}[tb]
 \begin{center}
 \includegraphics[width=0.48\textwidth]{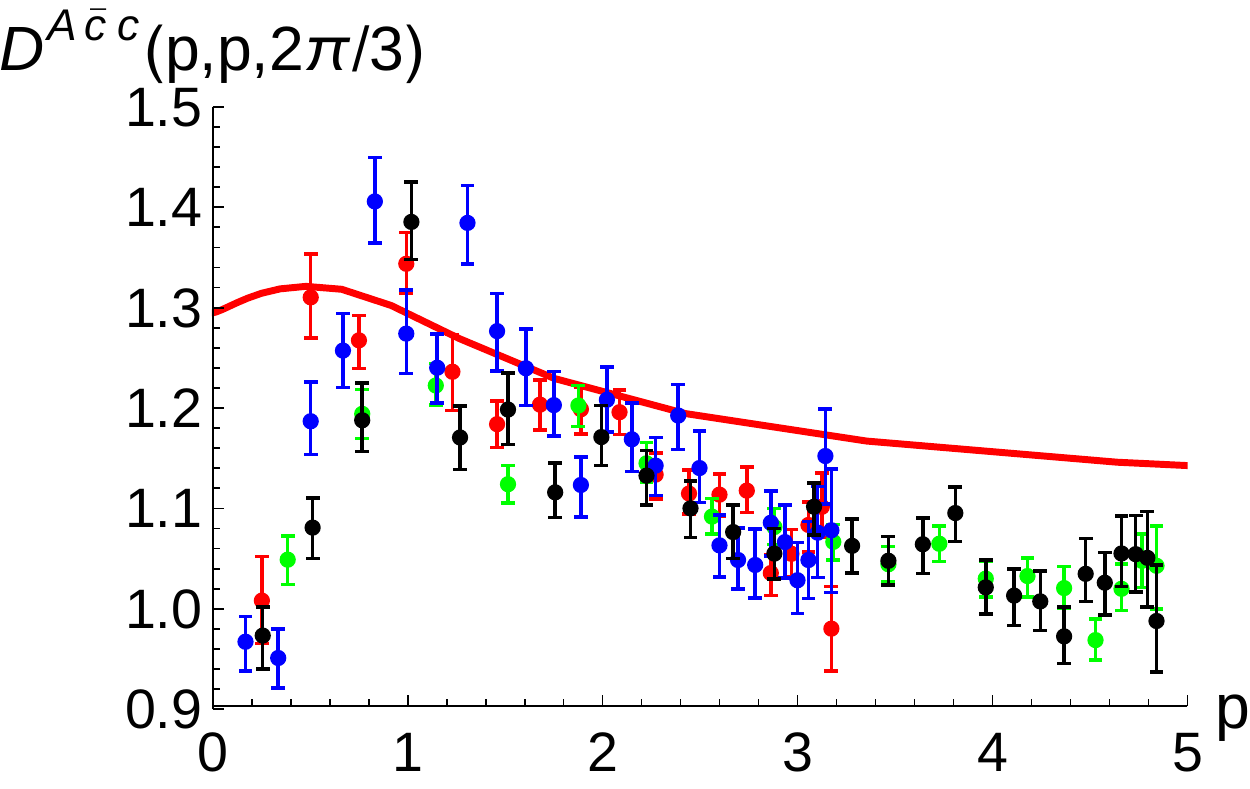}
 \hfill
 \includegraphics[width=0.48\textwidth]{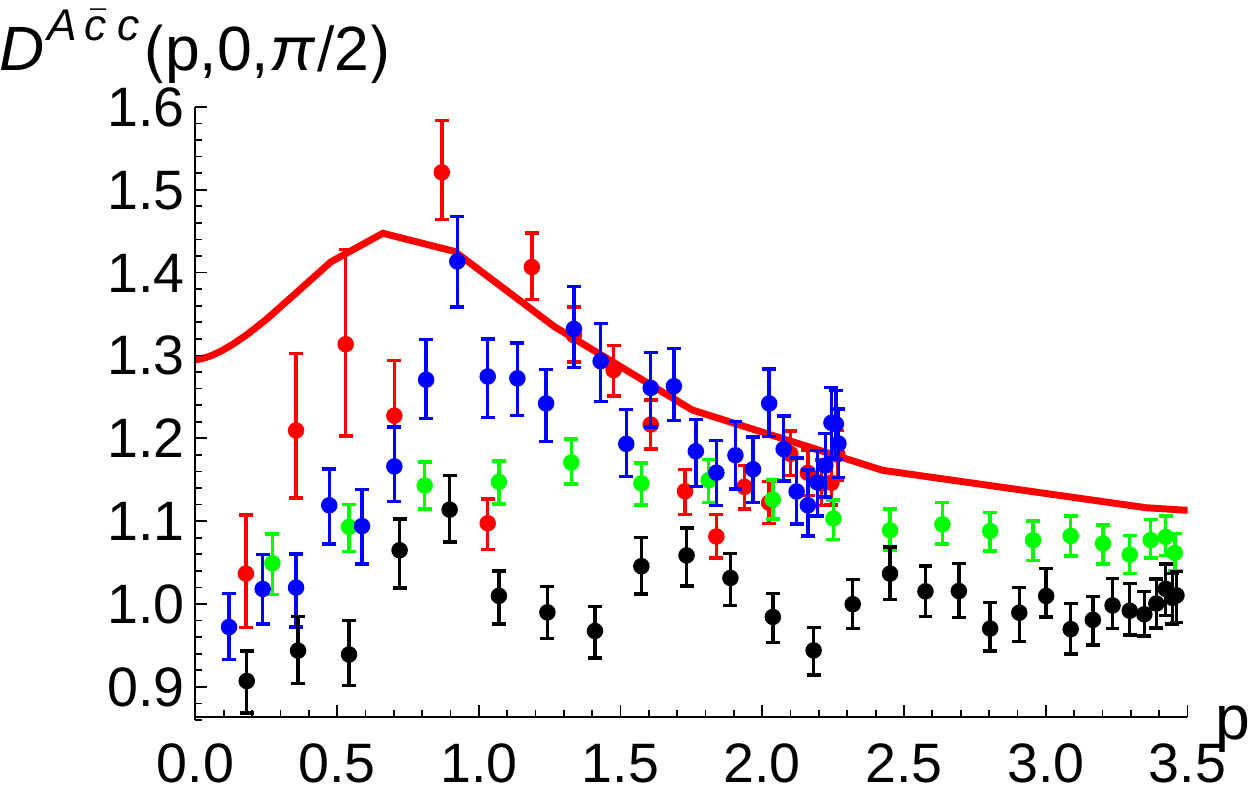}
 \caption{\label{fig:ghg_compL}Comparison to lattice results \cite{Cucchieri:2008qm} at the symmetric point (\textit{left}) and for vanishing gluon momentum (\textit{right}). Different colors correspond to different lattice sizes $N\in \{40,60\}$ and values for $\beta \in \{4.2,6\}$; see Ref.~\cite{Cucchieri:2008qm} for details.}
 \end{center}
\end{figure}

Figure \ref{fig:ghg_equalMoms} shows the ghost-gluon vertex dressing function
for equal momenta and different angles. There is only a slight dependence on the angle
between the momenta.
The selected values of the angle contain the two extreme points of parallel and anti-parallel
momenta ($\cos(\alpha)=-1$ and $1$) and the symmetric point ($\cos(\alpha)=-0.5$).
A comparison with lattice data from Landau gauge \cite{Cucchieri:2008qm} in three dimensions is shown in
\fref{fig:ghg_compL}. Qualitatively, the bump in the mid-momentum regime
is reproduced. However, quantitative agreement is not achieved. Most notably, the UV
regime is different. In Coulomb gauge the vertex possesses an anomalous dimension. In three-dimensional Landau gauge, on the other hand, 
it approaches the tree-level very quickly, because the gauge coupling is dimensionful in three dimensions and thus 
the vertex dressing must be suppressed as $1/p$ in the UV.
Lattice calculations \cite{Cucchieri:2008qm} and semi-perturbative DSE calculations \cite{Schleifenbaum:2004id} indeed show this behaviour.
At small momenta the lattice results drop back to $1$, while our results settle at a higher value.
This presumably reflects the two different type of solutions realized in Landau and Coulomb gauge,
respectively. Lattice calculations support the decoupling solution in Landau gauge \cite{Cucchieri:2007md,Cucchieri:2008fc,Sternbeck:2007ug,Bogolubsky:2009dc,Oliveira:2012eh}
but the scaling solution in Coulomb gauge \cite{Burgio:2008jr}. Thus it might not be appropriate to
compare the results obtained from the CRDSEs in Coulomb gauge with the lattice data for
the Landau gauge. The propagators of the two types of solutions differ mainly in the IR \cite{Fischer:2008uz} and the same is
expected for the corresponding ghost-gluon vertices as analogous investigations in Landau
gauge show \cite{Huber:2012kd}: The ghost-gluon vertex approaches the tree-level vertex
for the decoupling solution but receives a (finite) IR enhancement for the scaling solution.
This explains the difference between our results and lattice data in the IR.

\begin{figure}[tb]
 \begin{center}
 \includegraphics[width=0.48\textwidth]{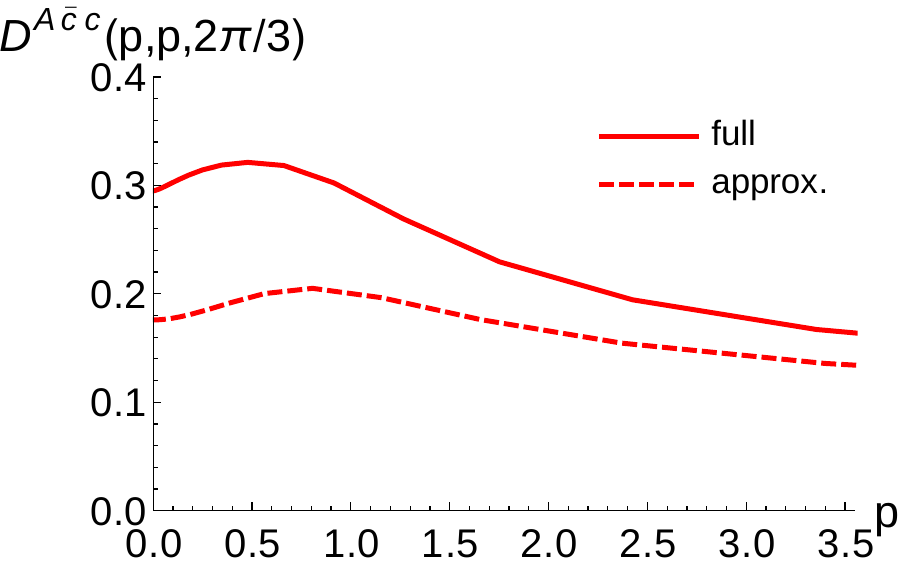} 
 \includegraphics[width=0.48\textwidth]{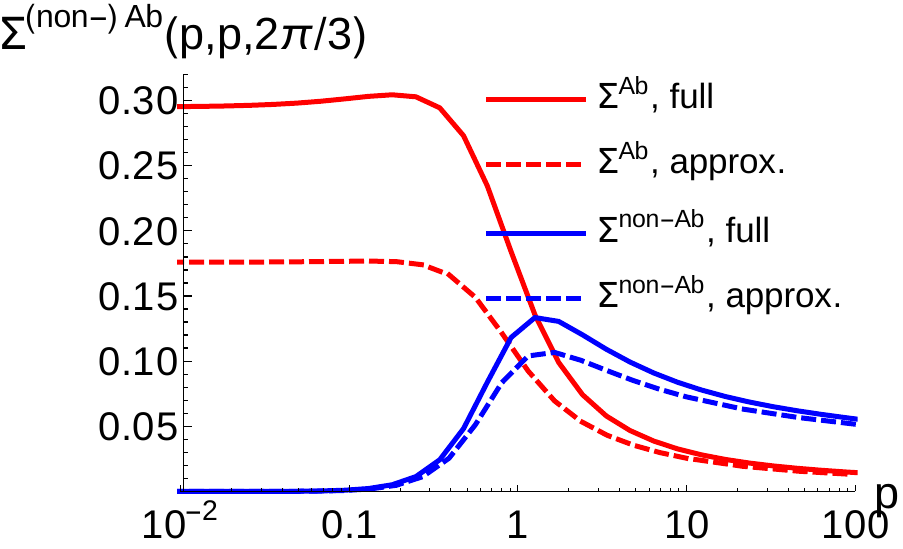}
 \caption{\label{fig:ghg_comp_bareV}Comparison of the full non-perturbative calculation
carried out in the present paper with the semi-perturbative calculation \cite{Campagnari:2011bk}
of the ghost-gluon vertex. \textit{Left}: Dressing function of the ghost-gluon vertex.
\textit{Right}: Contributions from the Abelian and non-Abelian diagrams.
The values for the non-Abelian diagram are larger than those in Fig.~2 of Ref.~\cite{Campagnari:2011bk},
where the coupling constant $g$ was factored out.}
 \end{center}
\end{figure}

In general, the results obtained for the ghost-gluon vertex are in accord with previous
investigations. 
For example, as anticipated for a scaling-type solution the ghost-gluon vertex stays finite in the IR \cite{Huber:2007kc,Campagnari:2011bk}. Also, 
it does not develop kinematic singularities in agreement with an IR analysis in three dimensions \cite{Alkofer:2008dt}.
Furthermore, the presently obtained dressing function of the ghost-gluon vertex has
qualitatively the same behavior as the one obtained in a
semi-perturbative calculation \cite{Campagnari:2011bk}
using full propagators (as in the present approach) but bare ghost-gluon vertices in
the loop diagrams of the CRDSEs.
In \fref{fig:ghg_comp_bareV} we compare the results of our full calculation with those of the semi-perturbative 
calculation of Ref.~\cite{Campagnari:2011bk}. While the non-Abelian diagram, which contains one
dressed ghost-gluon vertex, is not so much affected, the Abelian diagram gets much
more enhanced in the mid-momentum and IR regimes by using dressed vertices.

\subsection{Three-Gluon Vertex}
\label{sec:3g_results}

\begin{figure}[tb]
 \begin{center}
 \includegraphics[width=0.48\textwidth]{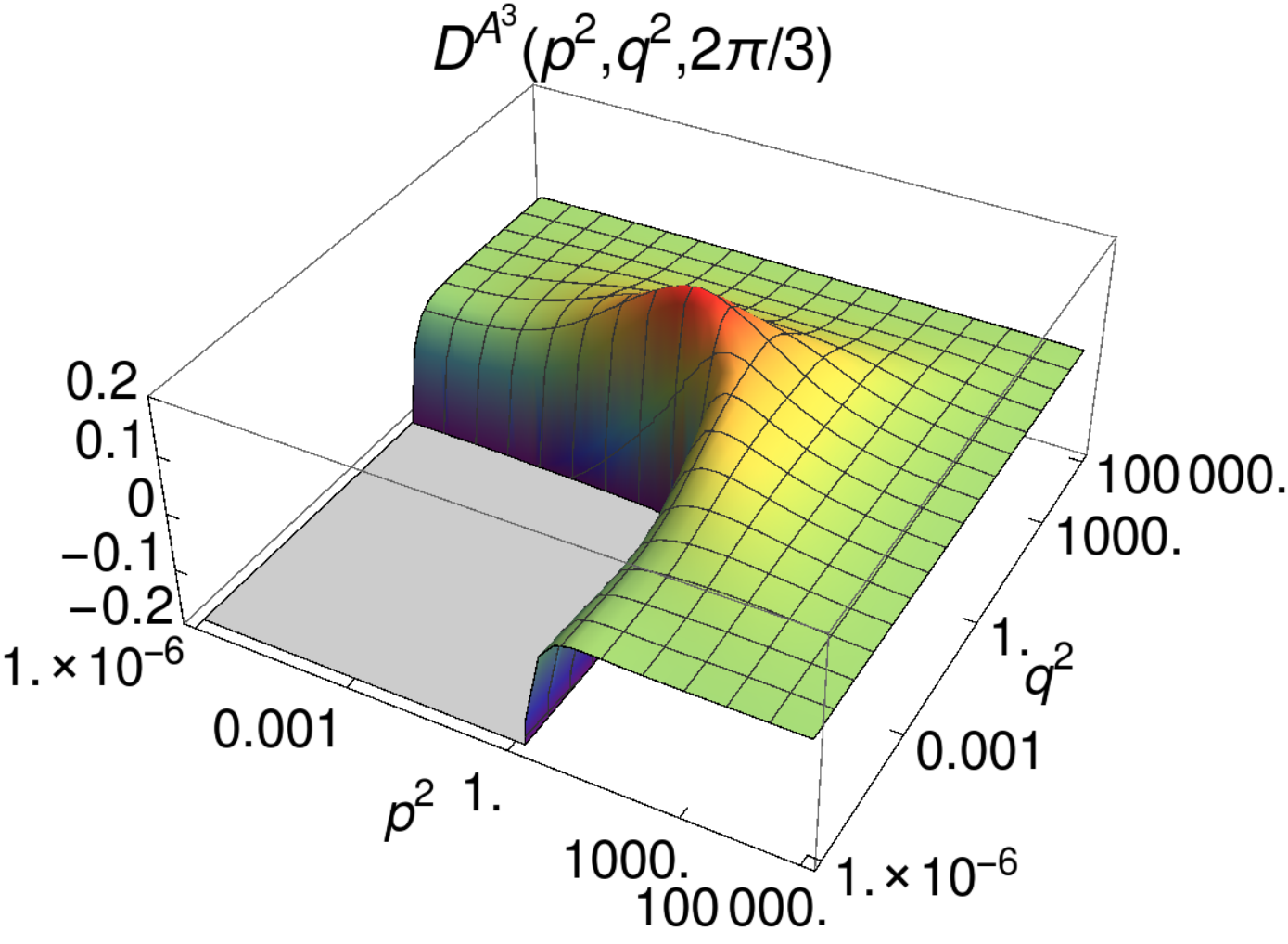}
 \includegraphics[width=0.48\textwidth]{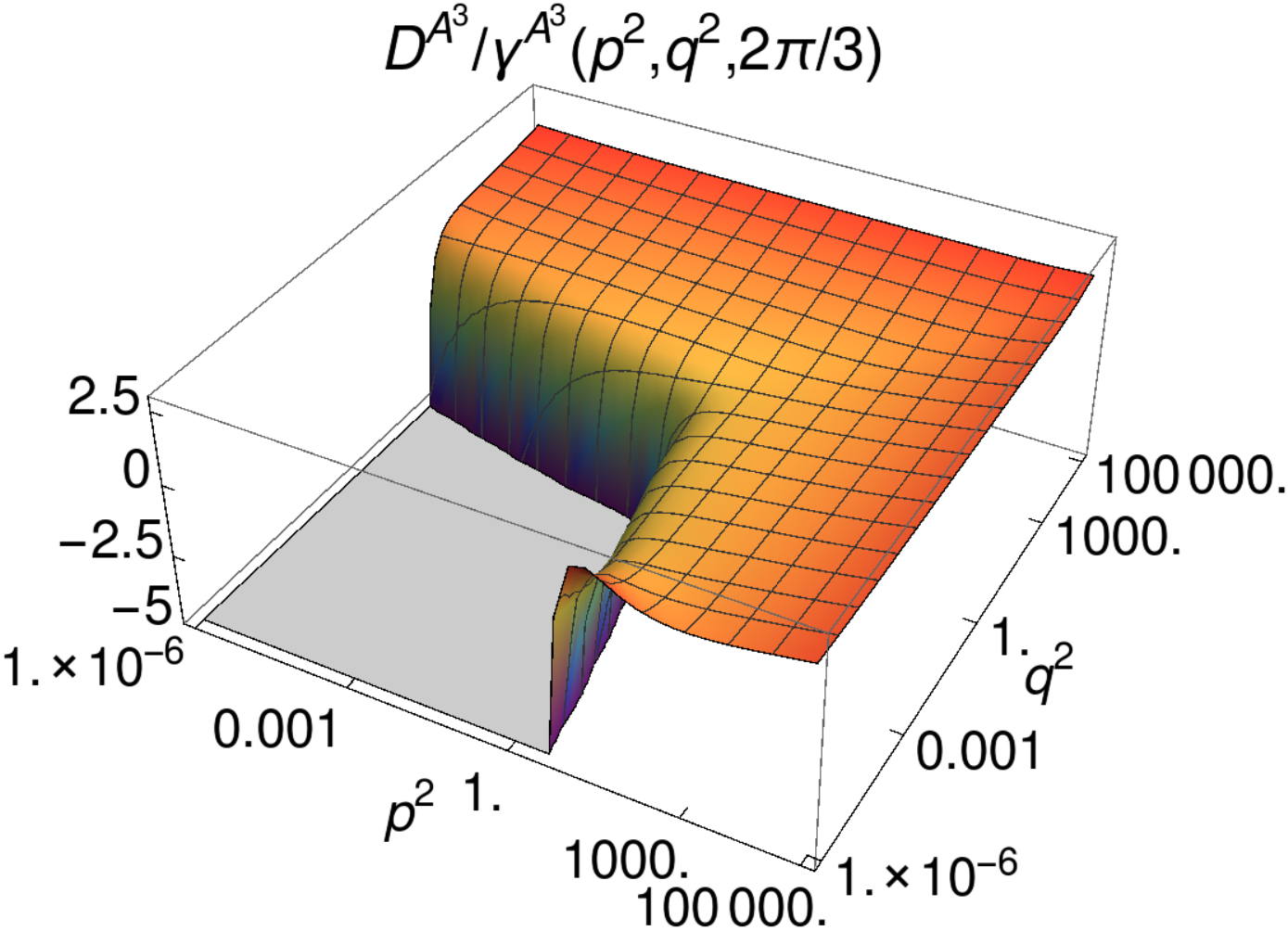}
 \caption{\label{fig:3g-3d}\textit{Left}: Dressing of the three-gluon vertex.
\textit{Right}: The ratio of three-gluon vertex to the variational kernel.
The deviation from Bose symmetry at the boundaries is a numerical artifact due to the smallness
of $\gamma_3$ which enhances small numerical errors considerably.}
 \end{center}
\end{figure}

\begin{figure}[tb]
 \begin{center}
 \includegraphics[width=0.48\textwidth]{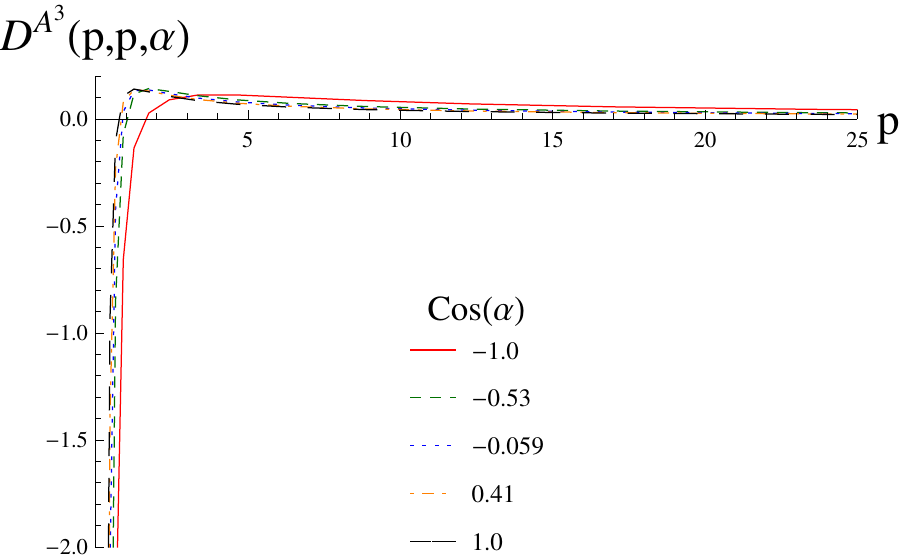}
 \includegraphics[width=0.48\textwidth]{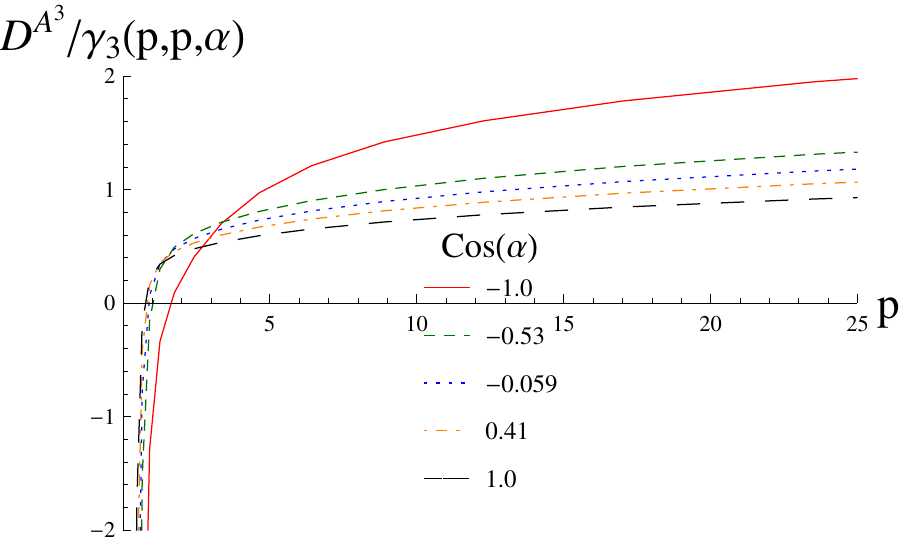}
 \caption{\label{fig:3g-equalMoms}\textit{Left}: Dressing function of the three-gluon vertex. \textit{Right}: Three-gluon vertex over variational kernel. Both plots for equal magnitude of two momenta with various angles between them.}
 \end{center}
\end{figure}

The dependence of the form factor of the three-gluon vertex on the magnitude of the external
momenta is shown in \fref{fig:3g-3d} for a fixed angle $\alpha$ [see the comment after \Eqref{v3proj}]
of roughly $2\pi/3$. The angle dependence of the dressing function is shown in \fref{fig:3g-equalMoms}. The selected
values of the angle contain the two extreme points of parallel and anti-parallel
momenta ($\cos(\alpha)=-1$ and $1$) and the symmetric point ($\cos(\alpha)=-0.5$).

\begin{figure}[tb]
 \begin{center}
 \includegraphics[width=0.48\textwidth]{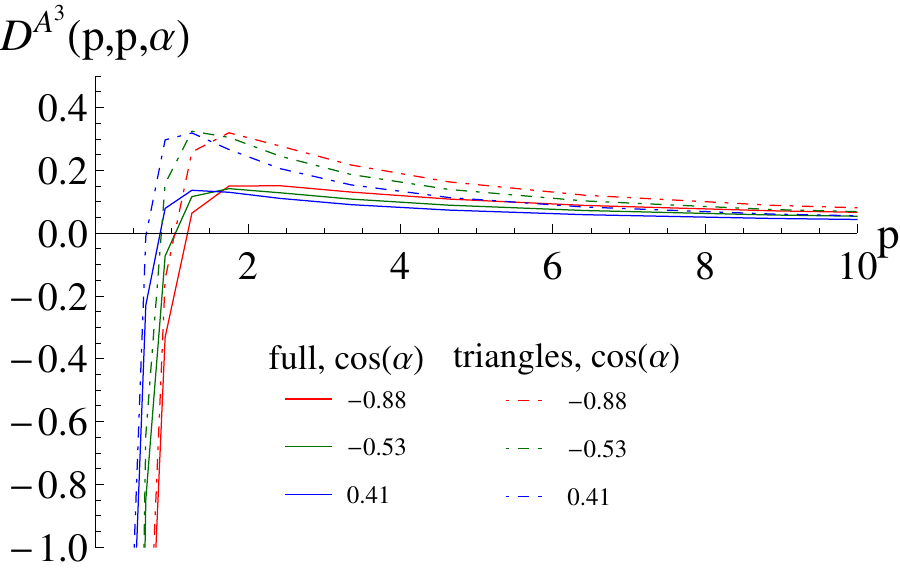}
 \includegraphics[width=0.48\textwidth]{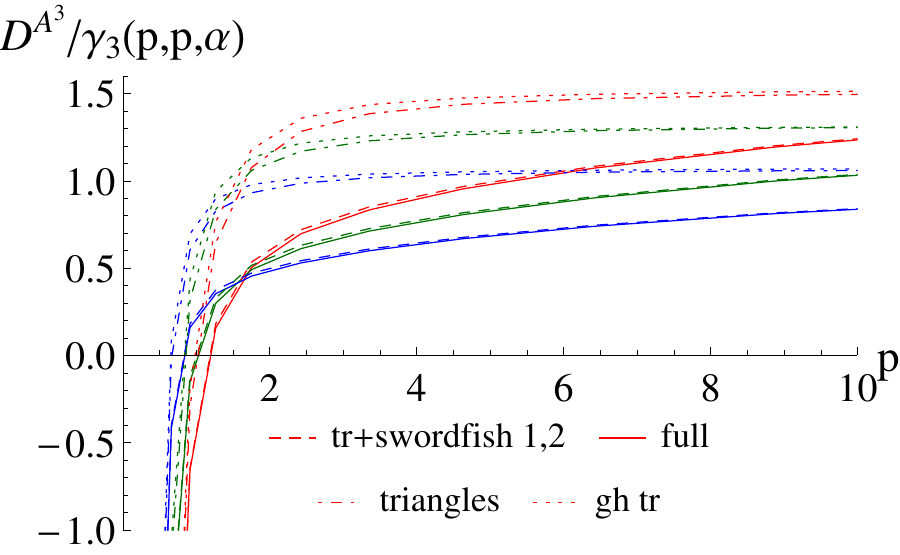}
 \caption{\label{fig:3g-compApprox}The three-gluon vertex from the full calculation (continuous line), from a simplified four-gluon kernel (dashed line), from a triangles-only calculation (dot-dashed line) and from a ghost-triangle-only calculation (dotted line). \textit{Left}: Three-gluon vertex dressing function. \textit{Right}: Ratio of the three-gluon vertex to the variational kernel. Colors correspond to the same angles as on the left panel.}
 \end{center}
\end{figure}

\begin{figure}[tb]
 \begin{center}
 \includegraphics[width=0.48\textwidth]{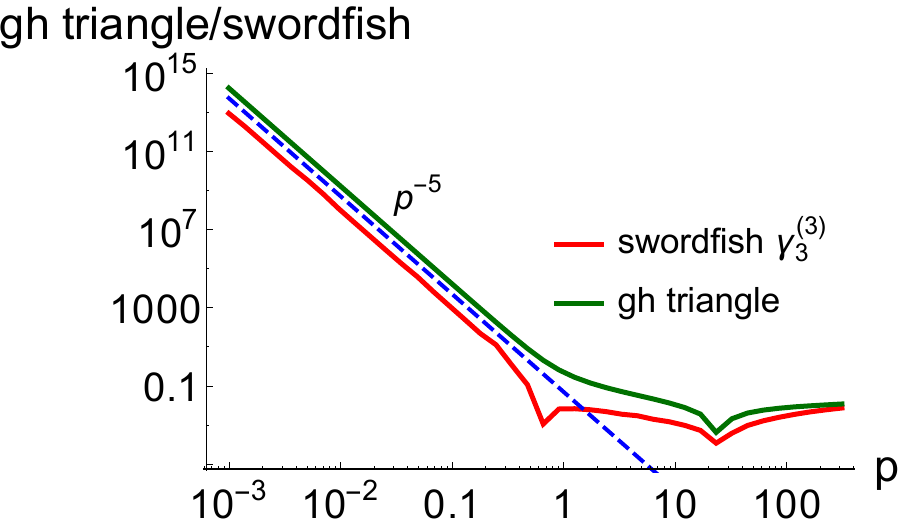}
 \caption{\label{fig:3g-compIR} IR behavior of ghost triangle (green, upper line) and the IR dominant part of the dynamic swordfish (red, lower line) projected as in \protect\Eqref{eq:selfenergy_sub_proj}. The dashed blue line is shown to illustrate the power law $p^{-5}$.}
 \end{center}
\end{figure}

A comparison between the full calculation and various approximative calculations is shown
in \fref{fig:3g-compApprox}.
The roughest approximation considered includes only the ghost triangle
diagram. We also show the result of the calculation where all triangle diagrams were
included. As can be seen it makes little difference to take all triangle diagrams
or only the ghost triangle. However, a comparison with the full calculation clearly shows
that neglecting the swordfish diagrams completely is too drastic an approximation.
Quite surprisingly, the calculation with a simplified
four-gluon kernel, where the contribution $\gamma_4^{(3)}$ [\Eqref{4gk-3}] which contains the Coulomb propagator
is neglected, can indeed reproduce the results from the full calculation rather well.
This is somewhat unexpected since, as mentioned in Sec.~\ref{sec:3g-DSE}, the term $\gamma_4^{(3)}$ [\Eqref{4gk-3}] of the
four-gluon kernel diverges like
$p^{-5}$ for small momenta. As a consequence, the swordfish diagram with a full three-gluon
vertex diverges in the IR with the same power as the ghost triangle. This is a peculiarity
of Coulomb gauge, which has no analogon in Landau gauge.
We have verified this explicitly and show a direct comparison in \fref{fig:3g-compIR}.
As can be seen, both diagrams (ghost triangle and the swordfish diagram with $\gamma_4^{(3)}$) diverge as $p^{-5}$ [$p^{-3}$ from the diagram and $p^{-2}$ from the projector \Eqref{eq:projector-3gv}].
Within the current truncation scheme, the magnitude of the IR dominant part of the
swordfish contribution is roughly 8\% of the ghost triangle; however, the sign is
opposite. As illustrated in \fref{fig:3g-compApprox}, neglecting this contribution leads
to very small deviations. Numerically, however, this contribution is one reason why,
compared to similar calculations in the Landau gauge \cite{Blum:2014gna,Huber:2012kd},
a higher precision is required here.
Taking into account the sum of the gluon energies in \Eqref{eq:DAAA} and the tensor structure
$T_3$ in \Eqref{v3proj}, the overall IR exponent of the full three-gluon vertex is $-3$,
in agreement with the results obtained in Refs.~\cite{Huber:2007kc,Campagnari:2010wc}.

\begin{figure}[tb]
 \begin{center}
 \includegraphics[width=0.48\textwidth]{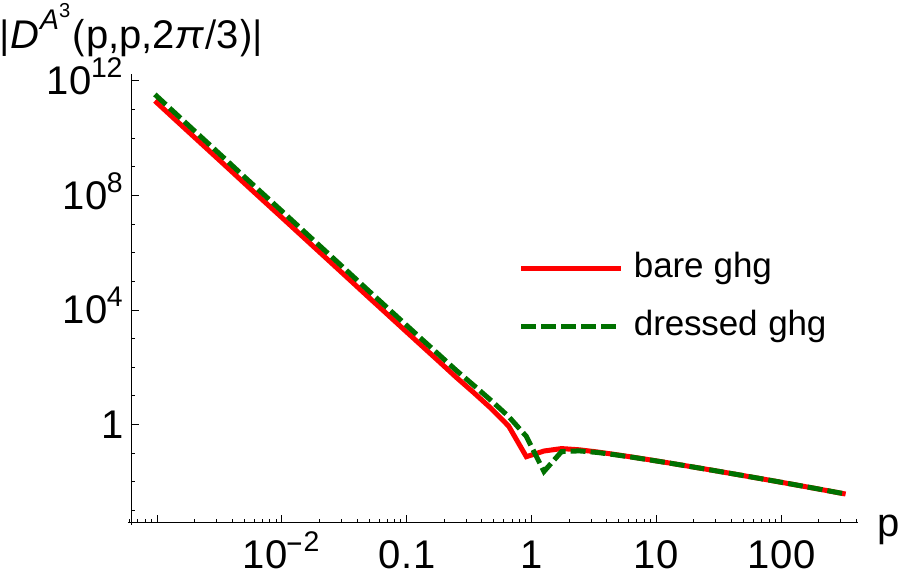}
 \hfill
 \includegraphics[width=0.48\textwidth]{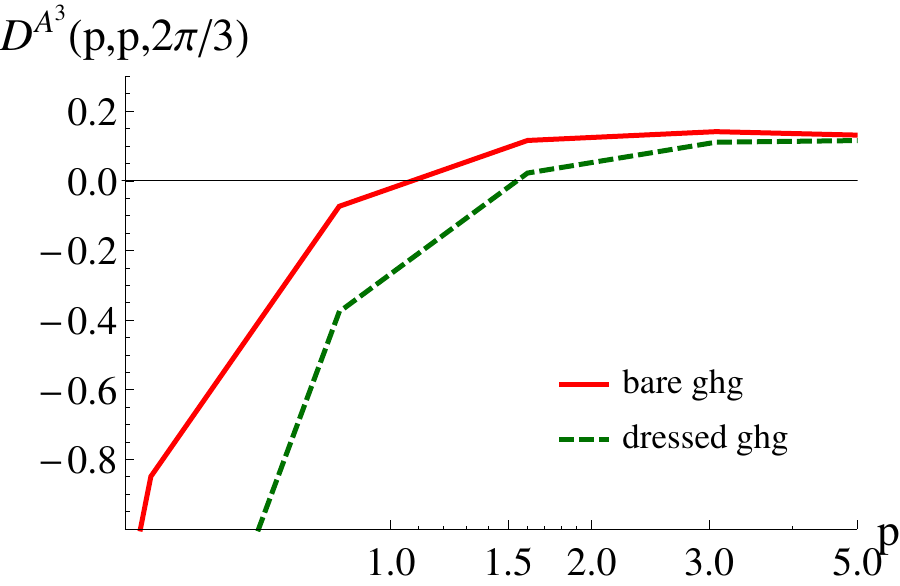}
 \caption{\label{fig:3g-compWithGhTrOnly} Three-gluon vertex dressing function at the symmetric point calculated with the full ghost-gluon vertex obtained here (green, dashed line) compared to results calculated with a bare ghost-gluon vertex (red, continuous line). The right panel shows the region around the zero crossing. }
 \end{center}
\end{figure}

In the results presented so far for the three-gluon vertex we have used a bare ghost-gluon vertex.
We have also solved the CRDSE~\eqref{3gvdsen} for the three-gluon vertex using the dressed
ghost-gluon vertex obtained in Sec.~\ref{sec:ghg_res}. The resulting dressing function
is shown in \fref{fig:3g-compWithGhTrOnly} and compared to that obtained with a
bare ghost-gluon vertex. As can be seen the difference is rather small. Both curves differ mainly in the IR,
where the coefficient of the power law is different. Also, the positions of the zeros of the
dressing functions differ.

\section{Summary}

We have numerically solved the CRDSEs for the ghost-gluon and three-gluon vertices self-consistently 
in a one-loop truncation using the ghost
and gluon propagators obtained previously with bare vertices as input. The ghost-gluon
vertex is somewhat infrared enhanced (but finite) and
drops gently with increasing momentum. It also shows little dependence on the angle between 
two momenta. Contrary to this, the dressing function of the three-gluon vertex is strongly infrared enhanced, in 
agreement with previous analytic analyses.
The Coulomb propagator enters the CRDSE for the three-gluon vertex through the four-gluon kernel:
while its contribution in the IR has the same power as the ghost loop, numerically it turns out to be almost negligible.
At higher momenta the gluon loop diagrams become important and dominate the quantitative behavior.
Furthermore, our numerical results show that in the calculation of the three-gluon vertex the dressing
of the ghost-gluon vertex can be ignored to good approximation.
The vertex dressings obtained in the present paper will serve as input in forthcoming studies within the variational approach to QCD in Coulomb gauge. 


\begin{acknowledgments}
M.H.~was supported by the Helmholtz International Center for FAIR within the LOEWE program of the State of Hesse and NAWI Graz.
D.C.~and H.R.~were supported by the Deutsche Forschungsgemeinschaft under contract No.~DFG-Re856/9-1.
\end{acknowledgments}

\appendix

\section{Kernels of the Ghost-Gluon Vertex Equation}

The kernels of the ghost-gluon vertex CRDSE are expressed in the following variables:
\begin{align}
\label{eq:sps}
 x&=\vp^2,& \quad y&=\vq^2, &\quad z&=\vk^2,&\quad \omega&=\vec{l}^2,\nnnl
 u&=\vp\cdot\vk,&\quad s&=\vk\cdot\vec{l}, & \quad v&=\vp\cdot\vec{l},
\end{align}
where $\vp$, $\vq$ and $\vk$ are external momenta and $\vec{l}$ is the loop momentum.
The arguments of the dressing functions are squared momenta.
The external momenta were chosen such that $\vp$ defines the 3-direction and $\vk$ lies in the 2-3-plane. 
The scalar products given in \Eqref{eq:sps} are then 
\[
 u=\sqrt{x z} \cos\varphi, \qquad
s=\sqrt{\omega z} (\cos \varphi \cos{\theta_2} + \sin{\varphi} \cos{\theta_1} \sin{\theta_2}), \qquad
v=\sqrt{x \omega} \cos \theta_2,
\]
where $\varphi$ is the angle between $\vp$ and $\vk$ and the integration angles are $\theta_1$ and $\theta_2$.

The self-energies of the ghost-gluon vertex, see \Eqref{eq:ghgDSE}, are given by
\begin{multline*}
  \Sigma^{\text{Ab}}(\vp,\vq;\vk)= g^2 N_c \int \frac{\d\omega \d\theta_1 \d\theta_2\,\sqrt{\omega}\sin(\theta_2)}{16\pi^3} \, K^{\text{ghg}}(\vp,\vk,\vec{l})\\
  \times\frac{d(\omega)d(2s+\omega+z)D^{\bar{c}cA}(x, \omega; -2 v + \omega + x])D^{\bar{c}cA}(2 s + \omega + z, 2 u + x + z; -2 v + \omega + x)}
  {4 \omega (-2 v + \omega + x) (-u^2 + x y) (2 s + \omega + z)\Omega(-2 v + \omega + x)},\\
\end{multline*}
\begin{multline*}
  \Sigma^{\text{non-Ab}}(\vp,\vq;\vk)= g^2 N_c \int \frac{d\omega d\theta_1 d\theta_2\,\sqrt{\omega}\sin(\theta_2)}{16\pi^3} \, L^{\text{ghg}}(\vp,\vk,\vec{l})\\
  \times\frac{d(2 v + \omega + x)D^{\bar{c}cA}(x, -2 v + \omega + x; \omega)D^{\bar{c}cA}(-2 v + w + x, 2 u + x + z; 2 s + w + z)}
  {2 \omega (-2 v + \omega + x) (-u^2 + x y) \Omega(\omega) \Omega(2 s + \omega + z) (\Omega(\omega) + \Omega(z) + \Omega(2 s + \omega + z))}.
\end{multline*}
for the equation with the gluon leg attached to the variational kernel [\Eqref{ggvdse1}] and
\begin{multline*}
  \Sigma^{\text{Ab}}(\vp,\vq;\vk)= g^2 N_c \int \frac{\d\omega \d\theta_1 \d\theta_2\,\sqrt{\omega}\sin(\theta_2)}{16\pi^3} \, K^{ghg}(\vp,\vk,\vec{l})\\
  \times\frac{d(\omega)d(2s+\omega+z)D^{\bar{c}cA}(\omega,2 s + \omega + z; z)D^{\bar{c}cA}(2 s + \omega + z,2 u + x + z; -2 v + \omega + x)}
  {4 \omega (-2 v + \omega + x) (-u^2 + x y) (2 s + \omega + z)\Omega(x+\omega-2v)},\\
\end{multline*}
\begin{multline*}
  \Sigma^{\text{non-Ab}}(\vp,\vq;\vk)= g^2 N_c \int \frac{d\omega d\theta_1 d\theta_2\,\sqrt{\omega}\sin(\theta_2)}{16\pi^3} \, L^{ghg}(\vp,\vk,\vec{l})\\
  \times\frac{d(-2 v + \omega + x)D^{A^3}(z, \omega, 2 s + \omega + z)D^{\bar{c}cA}(-2 v + \omega + x,2 u + x + z; 2 s + \omega + z)}
  {2 \omega (-2 v + \omega + x) (-u^2 + x y) \Omega(\omega) \Omega(2 s + \omega + z)}
\end{multline*}
for the equation with the anti-ghost leg attached to the bare vertex [\Eqref{ggvdse2}]. The explicit kernels read
\begin{align*}
K^{ghg}(\vp,\vk,\vec{l})={}& (v^2 + u (v - \omega) + s (v - x) - \omega x) (s u - v z),\\
L^{ghg}(\vp,\vk,\vec{l})={}& ((-u^2 + x y) (-s^2 (-2 v + \omega) + s v z + \omega (-v + \omega) z \\
& -u \omega (s + z)) + (s u - v z) (s^2 v + s (v^2 - 2 \omega x + v (\omega + z)) \\
& +\omega (v^2 + v z - x (\omega + z)) - u (s (-v + 2 \omega) + \omega (-v + \omega + z))).
\end{align*}


\bibliographystyle{h-physrev5}
\bibliography{biblio-spires}

\begin{thebibliography}{10}

\bibitem{Schutte:1985sd}
D.~Schutte,
\newblock Phys. Rev. {\bf D31}, 810 (1985).

\bibitem{Szczepaniak:2001rg}
A.~P. Szczepaniak and E.~S. Swanson,
\newblock Phys. Rev. {\bf D65}, 025012 (2001), arXiv:hep-ph/0107078.

\bibitem{Feuchter:2004mk}
C.~Feuchter and H.~Reinhardt,
\newblock Phys. Rev. {\bf D70}, 105021 (2004), arXiv:hep-th/0408236.

\bibitem{Epple:2006hv}
D.~Epple, H.~Reinhardt, and W.~Schleifenbaum,
\newblock Phys. Rev. {\bf D75}, 045011 (2007), arXiv:hep-th/0612241.

\bibitem{Reinhardt:2007wh}
H.~Reinhardt and D.~Epple,
\newblock Phys. Rev. {\bf D76}, 065015 (2007), arXiv:0706.0175.

\bibitem{Reinhardt:2008ek}
H.~Reinhardt,
\newblock Phys. Rev. Lett. {\bf 101}, 061602 (2008), arXiv:0803.0504.

\bibitem{Reinhardt:2011hq}
H.~Reinhardt, D.~Campagnari, and A.~Szczepaniak,
\newblock Phys. Rev. {\bf D84}, 045006 (2011), arXiv:1107.3389.

\bibitem{Heffner:2012sx}
J.~Heffner, H.~Reinhardt, and D.~R. Campagnari,
\newblock Phys. Rev. {\bf D85}, 125029 (2012), arXiv:1206.3936.

\bibitem{Reinhardt:2012qe}
H.~Reinhardt and J.~Heffner,
\newblock Phys. Lett. {\bf B718}, 672 (2012), arXiv:1210.1742.

\bibitem{Reinhardt:2013iia}
H.~Reinhardt and J.~Heffner,
\newblock Phys. Rev. {\bf D88}, 045024 (2013), arXiv:1304.2980.

\bibitem{Burgio:2008jr}
G.~Burgio, M.~Quandt, and H.~Reinhardt,
\newblock Phys. Rev. Lett. {\bf 102}, 032002 (2009), arXiv:0807.3291.

\bibitem{Campagnari:2010wc}
D.~R. Campagnari and H.~Reinhardt,
\newblock Phys. Rev. {\bf D82}, 105021 (2010), arXiv:1009.4599.

\bibitem{Reinhardt:2008ij}
H.~Reinhardt and W.~Schleifenbaum,
\newblock Annals Phys. {\bf 324}, 735 (2009), arXiv:0809.1764.

\bibitem{Schleifenbaum:2006bq}
W.~Schleifenbaum, M.~Leder, and H.~Reinhardt,
\newblock Phys. Rev. {\bf D73}, 125019 (2006), arXiv:hep-th/0605115.

\bibitem{Eichmann:2014xya}
G.~Eichmann, R.~Williams, R.~Alkofer, and M.~Vujinovic,
\newblock Phys.Rev. {\bf D89}, 105014 (2014), arXiv:1402.1365.

\bibitem{Blum:2014gna}
A.~Blum, M.~Q. Huber, M.~Mitter, and L.~von Smekal,
\newblock Phys. Rev. D {\bf 89}, 061703(R) (2014), arXiv:1401.0713.

\bibitem{Huber:2007kc}
M.~Q. Huber, R.~Alkofer, C.~S. Fischer, and K.~Schwenzer,
\newblock Phys. Lett. {\bf B659}, 434 (2008), arXiv:0705.3809.

\bibitem{Wolfram:2004}
S.~Wolfram,
\newblock {\em The Mathematica Book} (Wolfram Media and Cambridge University
  Press, 2004).

\bibitem{Alkofer:2008nt}
R.~Alkofer, M.~Q. Huber, and K.~Schwenzer,
\newblock Comput. Phys. Commun. {\bf 180}, 965 (2009), arXiv:0808.2939.

\bibitem{Huber:2011qr}
M.~Q. Huber and J.~Braun,
\newblock Comput. Phys. Commun. {\bf 183}, 1290 (2012), arXiv:1102.5307.

\bibitem{Schleifenbaum:2004dt}
W.~Schleifenbaum,
\newblock diploma thesis, Eberhard-Karls-Universit\"at zu T\"ubingen, 2004.

\bibitem{Huber:2011xc}
M.~Q. Huber and M.~Mitter,
\newblock Comput. Phys. Commun. {\bf 183}, 2441 (2012), arXiv:1112.5622.

\bibitem{Cucchieri:2008qm}
A.~Cucchieri, A.~Maas, and T.~Mendes,
\newblock Phys. Rev. {\bf D77}, 094510 (2008), arXiv:0803.1798.

\bibitem{Schleifenbaum:2004id}
W.~Schleifenbaum, A.~Maas, J.~Wambach, and R.~Alkofer,
\newblock Phys.Rev. {\bf D72}, 014017 (2005), arXiv:hep-ph/0411052.

\bibitem{Cucchieri:2007md}
A.~Cucchieri and T.~Mendes,
\newblock PoS {\bf LAT2007}, 297 (2007), arXiv:0710.0412.

\bibitem{Cucchieri:2008fc}
A.~Cucchieri and T.~Mendes,
\newblock Phys. Rev. {\bf D78}, 094503 (2008), arXiv:0804.2371.

\bibitem{Sternbeck:2007ug}
A.~Sternbeck, L.~von Smekal, D.~Leinweber, and A.~Williams,
\newblock PoS {\bf LAT2007}, 340 (2007), arXiv:0710.1982.

\bibitem{Bogolubsky:2009dc}
I.~L. Bogolubsky, E.~M. Ilgenfritz, M.~M\"uller-Preussker, and A.~Sternbeck,
\newblock Phys. Lett. {\bf B676}, 69 (2009), arXiv:0901.0736.

\bibitem{Oliveira:2012eh}
O.~Oliveira and P.~J. Silva,
\newblock Phys.Rev. {\bf D86}, 114513 (2012), arXiv:1207.3029.

\bibitem{Fischer:2008uz}
C.~S. Fischer, A.~Maas, and J.~M. Pawlowski,
\newblock Annals Phys. {\bf 324}, 2408 (2009), arXiv:0810.1987.

\bibitem{Huber:2012kd}
M.~Q. Huber and L.~von Smekal,
\newblock JHEP {\bf 1304}, 149 (2013), arXiv:1211.6092.

\bibitem{Campagnari:2011bk}
D.~R. Campagnari and H.~Reinhardt,
\newblock Phys. Lett. {\bf B707}, 216 (2012), arXiv:1111.5476.

\bibitem{Alkofer:2008dt}
R.~Alkofer, M.~Q. Huber, and K.~Schwenzer,
\newblock Eur. Phys. J. {\bf C62}, 761 (2009), arXiv:0812.4045.

\end{thebibliography}

\end{document}